\documentclass[twocolumn,aip,jcp,reprint,floatfix]{revtex4-1}

\pdfoutput=1

 \usepackage{color}

\usepackage{graphicx} 
\usepackage{amssymb,amsmath} 
\usepackage{verbatim} 
\usepackage{mathrsfs}
\usepackage{epstopdf}

\usepackage[normalem]{ulem}

\begin{document}	

\newcommand{\sanseb}{Nano-Bio Spectroscopy Group and ETSF Scientific Development Center, University of the Basque Country UPV/EHU, Avenida de Tolosa 72, 20018 San Sebastian, Spain}
\newcommand{\ifgw}{Instituto de F\'isica ''Gleb Wataghin'', Universidade Estadual de Campinas, 13083-970, Campinas, S\~ao Paulo, Brazil }
\newcommand{\csic}{Centro de F\'isica de Materiales CSIC-UPV/EHU-MPC and DIPC, Av.~Tolosa 72, E-20018 San Sebasti\'an, Spain}
\newcommand{\fhi}{Fritz-Haber-Institut der Max-Planck-Gesellschaft,Faradayweg 4-6, D-14195 Berlin, Germany}

\title{Simulating pump-probe photo-electron and absorption spectroscopy on the attosecond 
			 time-scale with time-dependent density-functional theory}

\date{\today}
\author{Umberto De Giovannini}\email{umberto.degiovannini@ehu.es} \affiliation{\sanseb}
\author{Gustavo Brunetto}\affiliation{\sanseb}\affiliation{\ifgw} 
\author{Alberto Castro}\email{acastro@bifi.es}\affiliation{ARAID Foundation -- Institute for Biocomputation and Physics of Complex Systems
                               (University of Zaragoza), Zaragoza, Spain}
\author{Jessica Walkenhorst}\affiliation{\sanseb}
\author{Angel Rubio}\email{angel.rubio@ehu.es}\affiliation{\sanseb}\affiliation{\csic}\affiliation{\fhi}

\begin{abstract}
Molecular absorption and photo-electron spectra can be efficiently predicted with
real-time time-dependent density-functional theory (TDDFT). We show here how these
techniques can be easily extended to study time-resolved pump-probe experiments
in which a system response (absorption or electron emission) to a probe
pulse, is measured in an excited state. This simulation tool helps to interpret 
the fast evolving attosecond time-resolved spectroscopic experiments, where
the electronic motion must be followed at its natural time-scale. We show how
the extra degrees of freedom (pump pulse duration, intensity, frequency, and
time-delay), which are absent in a conventional steady state experiment, provide additional
information about electronic structure and dynamics that improve a system characterization.
As an extension of this approach, time-dependent 2D spectroscopies can also be simulated,  
in principle, for large-scale structures and extended systems.
\end{abstract}
\maketitle

\section{Introduction}


Pump-probe experiments are the preferred technique to study the
dynamical behaviour of atoms and molecules: the dynamics triggered by
the pump pulse can be monitored by the time-dependent reaction of the
system to the probe pulse, a reaction that can be measured in terms of,
for example, the absorption of the pulse intensity, or of the
emission of electrons.  \cite{krehl1995note} The time resolution of
these experiments is mainly limited by the duration of the pulses --
although it is also limited by the ability of the experimenter to
ascertain their relative time delay and shape. In order to precisely fix this
delay, the two pulses are coherently synchronized -- in fact, they
have the same origin, or one of them is used to generate the other --,
so that the delay is gauged by an optical path
difference.\cite{abraham1899note} The electron dynamics has a natural
time-scale in the range of attoseconds, and therefore could not be
studied by pump-probe spectroscopy until the advent of
attosecond-pulse laser sources one decade
ago.\cite{Hentschel2001,Kienberger2004} Nowadays time-resolved
spectroscopy can be utilized to monitor electron dynamics in real
time, giving birth to the field of attosecond physics.
\cite{RevModPhys.81.163,0953-4075-39-1-R01}

A wealth of possibilities exists, depending on the frequencies,
durations and intensities of the two pulses. A common set-up in
attosecond physics employs a XUV attosecond pulse and the relatively
more intense NIR or visible longer (few femtoseconds) pulse used for
its generation. Combining two XUV attosecond pulses is in principle
possible (and has been theoretically
analysed\cite{PhysRevA.85.013415}), but unfortunately the low
outputs of current XUV attosecond pulses lead to much too weak signals. 
Another choice to make is the final observable, i.e. what
kind of system reaction is to be measured as a function of the time
delay. In this work we focus on two common choices. One, observing the
emission of electrons (their energies, angular distribution, or total
yield) from the pumped system due to the probe pulse. This can be
called time-resolved photo-electron spectroscopy (TRPES). Two,
observing the optical absorption of the probe signal, which can
be called time resolved absorption spectroscopy, or transient
absorption spectroscopy (TAS).

Both techniques can of course be used to look at longer time
resolutions -- and there is already a substantial body of literature
describing such experiments.  If we look at molecular reaction on the
scale of tens or hundreds of femtoseconds, the atomic structure will
have time to re-arrange. These techniques are thus mainly
employed to observe modification, creation, or destruction of
bonds, a field now named \emph{femtochemistry}.\cite{zewail2000}

TAS, for example, has been succesfully employed to \emph{watch} the
first photo-synthetic events in cholorophylls and
carotenoids,\cite{berera2009} that transform the energy gained by
light absorption into molecular rearrangements. A review describing
the essentials of this technique can be found in
Ref.~\onlinecite{foggi2001}.  Note, however, that in addition to
following chemical reactions, femtosecond-long pulses may also be used
for example for characterizing the final electronic quantum state of
ionized atoms.\cite{Loh:2007ec}

In TRPES, the probe pulse generates free electrons through
photo-ionization, and one measures their energy or angular
distribution as a function of time; if this time is in the femtosecond
scale one can follow molecular dynamics in the gas phase, as
demonstrated already in the mid 1990s,\cite{Baumert199329,assion1996}
although this technique had already been employed to follow electronic
dynamics on surfaces.\cite{haight1985} 
This methodology is well documented in recent
articles.\cite{stolow2003,stolow2004,bauer2005,suzuki2006,Wu:2011cv}

If the goal --as in this work-- is to study the electronic dynamics
\emph{only}, disentangling them from the vibronic degrees of freedom,
then one must move down these spectroscopic methods to the attosecond
regime.\cite{pfeifer2008} In this regime, both TAS and TRPES have
recently been demonstrated. Regarding TAS, we may cite as
a prototypical example the recent experiment of Holler \emph{et
  al}\cite{holler2011} where, the transient
absorption of an attosecond pulse train (created by high harmonic
generation) by a Helium gas target, was studied in the presence of an
intense IR pulse. The absorption was observed to oscillate as a
function of the time-delay of pump and probe. Another example is the
real-time observation of valence electron motion reported by
Goulielmakis {\em et al}.\cite{Goulielmakis:2010br}

Several cases of use of TRPES with attosecond pulses have also
been recently reported. For example, Uiberacker {\em et
  al.}\cite{Uiberacker2007} could observe in real time the light
induced electron tunneling provoked by a strong NIR pulse,
demonstrating how this electron tunneling can be used to probe short
lived electronic states.  Smirnova {\em et al}\cite{smirnova2006} also
studied the ionization of an atom by an attosecond XUV pulse in the
presence of an intense laser pulse, as a function of the time delay
between both.  Jonnsson {\em et al}\cite{PhysRevLett.99.233001}
employed attosecond pulse trains and a Helium target, and attosecond
photoelectron spectroscopy was also demonstrated to yield useful
information for condensed matter systems.\cite{Cavalieri2007}

All these advances demand an appropriate theoretical modeling.  The
use of more than one pulse of light intrinsically requires to go
beyond any ``linear spectroscopy'' technique -- although if the pulses
are weak a perturbative treatment may still be in order. This
non-linear behavior, provides much more information about the
system at the cost of a increasingly difficult analysis. The use of two
(or more) coherent pulses of light, with fine control over their shape
(sometimes called a ``multidimensional analysis''), permits a deeper 
characterization. This fact was already acknowledged in the
field of nuclear magnetic resonance, or later in femtochemistry -- see
for example Refs.~\onlinecite{mukamel2004} and
\onlinecite{Pollard:1990up} for theoretical treatments of these cases.

A recent theoretical analysis of attosecond TAS based on perturbation
theory was given by Baggesen {\em et al}.~\cite{PhysRevA.85.013415}
Gaarde {\em et al}~\cite{gaarde2011} presented a study for
relatively weak pumping IR pulses in combination with XUV ultrafast
probes, for Helium targets and based on the single active electron
approximation (SAE). 
Very recently, the experimented reported by Ott {\em et al.}\cite{arxiv1205.0519},
in which the ultrafast TAS of Helium displayed characteristic beyond-SAE features,
was theoretically analyzed in Ref.~\onlinecite{arxiv1211.2566v1}, utilizing an exact
solution of the time-dependent Schr{\"{o}}dinger equation, that cannot however
be easily extended to larger systems.
Finally, the above-mentioned experiment of Goulielmalkis
{\em et al}\cite{Goulielmakis:2010br} was analized with the model
described in Ref.~\onlinecite{PhysRevA.83.033405}, which treated the pump
IR pulse non-perturbatively.

Indeed, it would be desirable to analyze these processes with a
non-perturbative theory (since at least one of the pulses is usually
very intense), which at the same time is capable of going beyond the
SAE and accounting for many-electron interaction effects.  This last
fact is relevant since the attosecond time resolution obtained in this
type of experiments is able to unveil the fast dynamical
electron-electron interaction effects. The SAE, which essentially
assumes that only one electron actively responds to the laser pulse,
has been successfully used to interpret many strong-field
processes. However, its range of validity is limited, and roughly
speaking it is expected to fail whenever the energies of multielectron
excitations become comparable to the laser frequencies or the single
electron excitations.\cite{lezius2001}

Time-dependent density functional theory (TDDFT)~\cite{tddft} in principle
meets all requirements: it may be used non-perturbatively,
includes the electron-electron interaction and can handle out-of-equilibrium situations. 
It has been routinely used in the past decades to study the electron dynamics in condensed matter
\emph{in equilibrium}. By this we mean that, usually, one computes the
linear or non-linear response properties of systems in the ground
state (or at thermal equlibrium). In pump-probe experiments, however,
one must compute the response of a system that is being driven out of equilibrium by an
initial pulse. In this work, we will explore the usability of TDDFT
for this purpose, and show how, at least for the two cases of TAS and
TRPES, the extension is straightforward.

\section{Theory}

Density-functional theory~\cite{kohn1999} establishes a one-to-one
correspondence between the ground-state density and the external potential
of a many-electron system. This implies that any system property is, in
principle, a ground-state density functional. The computation of the
ground-state density usually follows the Kohn-Sham (KS)\cite{kohn1965}
scheme, in which one utilizes a fictitious system of non-interacting
electrons that has the same ground state density. For excited states, 
properties, however, or in order to simulate the behavior of the
system in time-dependent external fields, 
one must use its time-dependent version, TDDFT.\cite{Runge:1984us,tddft,cptddtf:2011}

In the case of TDDFT, a one-to-one correspondence also exists between the
time-dependent densities and potentials. 
One also uses an auxiliary fictitious system of non-interacting electrons that produces the same
time-dependent density. This substitution is the source of the great
computational simplification, since a non-interacting system of
electrons can in general be represented by a single Slater determinant
formed by a set of ``Kohn-Sham'' orbitals, $\varphi_i$
($i=1,\dots,N/2$). We will assume a spin-compensated system of $N$
electrons doubly occupying $N/2$ spatial orbitals. If the real system
is irradiated with an external field characterized by a scalar
potential $v(\vec{r},t)$ (the extension to vector potentials is also possible), 
the ``time-dependent Kohn-Sham'' (TDKS) equations that characterize the evolution of the
fictitious system are (atomic units will be used hereafter):
\begin{eqnarray}
{\rm i}\frac{\partial}{\partial t}\varphi_i(\vec{r},t) & = & -\frac{1}{2}\nabla^2 \varphi_i(\vec{r},t)
+ v_{\rm KS}[n](\vec{r},t)\varphi_i(\vec{r},t)\,,
\\
n(\vec{r},t) & = & 2\sum_{i=1}^{N/2}\vert\varphi_i(\vec{r},t)\vert^2\,.
\end{eqnarray}
The time-dependent density $n(\vec{r},t)$ is the central object, and is 
identical for the real and for the KS systems. The KS potential $v_{\rm KS}$
is a functional of this density, and is defined as:
\begin{equation}
v_{\rm KS}[n](\vec{r},t) = v_0(\vec{r}) + v(\vec{r},t) + v_{\rm H}[n](\vec{r},t) + v_{\rm xc}[n](\vec{r},t)\,,
\end{equation}
where the Hartree potential $v_{\rm H}$ corresponds to a classical electrostatic term
\begin{equation}
v_{\rm H}[n](\vec{r},t) = \int\!\!{\rm d}^3r'\;\frac{n(\vec{r},t)}{\vert \vec{r}'-\vec{r}\vert}\,,
\end{equation}
$v_0(\vec{r})$ is the static external potential that characterizes the system
in its ground state (in a molecule, originated by a set of nuclei),
and the ``exchange and correlation'' potential is $v_{\rm xc}[n]$.
The exchange correlation potential is also a functional of the density and accounts for all the intricate many-electron effects.
It is in practice unknown and must be approximated.\cite{tddft,Marques20122272}

The TDKS equations can be utilized, either directly or in appropriately
transformed manner, to compute the response of a many-electron system
to a perturbation, weak or strong. In the perturbative regime, ideally
one wishes to obtain the response functions [(hyper)-polarizabilities,
optical and magnetic susceptibilities, \dots], since (i) these objects then permit to
predict any reaction in the appropriate order, and (ii) experiments
typically provide spectra that are directly related to the response
functions -- e.g. the optical absorption cross section of a gas is
proportional to the imaginary part of the dipole-dipole molecular
polarizability. 
In contrast, in the strong-field regime, where perturbative treatments become cumbersome, 
one normally computes the particular response of
the system to the perturbation of interest by directly propagating 
the TDKS equations in real time.

The vast majority of TDDFT applications have addressed the first-order
response of the ground-state system to weak electric fields -- which
can provide the absorption spectrum, the optically-allowed excitation
energies and oscillator strengths, etc. This can be performed by
linearizing the TDKS equations in the frequency domain and casting the
result into matrix-eigenvalue form, or by propagating the same
equations in real time applying a sufficiently weak dipole
perturbation. In any case, the response function computed in this
manner will be that of the ground state. If we want to analyze a TAS
experiment, the objective is to obtain the response of the excited
states that are visited by the system as it is driven by the pump
pulse (i.e. the response function of a system out 
of equilibrium). 
This extension will be treated in Section~\ref{subsection:tas}.

Likewise, TDDFT can be used to compute strong field non-linear
photo-electron spectra of atoms and molecules, for example with the method recently
developed by some of us.\cite{UDeGiovannini:2012hy} These spectra,
however, are also characteristic of the ground state, although,
as it will be shown in Section~\ref{subsection:pes}, the methodology
can be easily extended to tackle the pump-probe case
(time-resolved photoelectron spectroscopy~\cite{Wu:2011cv}).

\subsection{Attosecond transient absorption spectroscopy}
\label{subsection:tas}

When an electromagnetic pulse passes through a gas sample, the molecules
polarize, and this polarization modifies the otherwise free
propagation of light -- one of the consequences being the partial
absorption of it. In a dilute gas, assuming the electric dipole
approximation and a sufficiently weak pulse, the dipole-dipole linear
dynamical polarizability entirely determines the polarization of the
medium, and therefore the amount of absorption. This is usually
understood \emph{at equilibrium}: the gas is formed by molecules at
thermal equilibrium (perhaps at sufficiently low temperature so that
they all can be considered to be at their ground state), and the only
light pulse present is that whose absorption we want to measure.

In the pump-probe situation discussed here, however, one wishes to
compute the absorption of a probe pulse by a set of molecules that
is also irradiated by a pump, either simultaneously or with a given
delay. The task is therefore to compute the response of the electric
dipole with and without the probe pulse -- the difference being the
excess of polarization, responsible for the absorption of the probe. 
We will assume, as it is often the case, that the pump pulse is
intense, whereas the probe is weak and can be treated in first order
perturbation theory.

This situation is amenable to a generalized definition of response
functions, such as the one given in the appendix of
Ref.~\onlinecite{castro2011} and also discussed in detail in Ref.~\onlinecite{Mukamel:1999}. 
We will review here this definition,
adapting it to the pump-probe situation. Let us depart from a
Hamiltonian in the form (we only treat the electric part neglecting 
the magnetic term of the electromagnetic field):
\begin{equation} 
  \hat{H}_0 [\mathcal{E}](t) = \hat{\mathcal{H}} + \mathcal{E}(t) \hat{V}
  \label{eq:hamiltonian-time-dependent}
\end{equation}
where $\hat{\mathcal{H}}$ is the static Hamiltonian that defines the system
itself, and $\mathcal{E}(t)\hat{V}$ is the coupling to the ``pump'' laser 
pulse. 
This is the ``unperturbed'' Hamiltonian, that contains only the pump
pulse; the full Hamiltonian results of the addition of the probe pulse
$f(t)\hat{V}$:
\begin{equation}
  \hat{H}(t) = \hat{\mathcal{H}} + \mathcal{E}(t) \hat{V} + f(t)\hat{V}\,.
  \label{eq:hamiltonian-full}
\end{equation}
The evolution of the system is given by:
\begin{equation}
{\rm i}\frac{\partial}{\partial t} \hat{\rho}(t) = \left[\hat{H}(t),\hat{\rho}(t)\right]\,,
\end{equation}
and initially ($t = t_0$, some time before the arrival of both pump or probe), the system is at equilibrium:
\begin{equation}
  \left[ \hat{\mathcal{H}}, \hat{\rho}_{eq} \right] = 0 \,.
  \label{eq:equilibrium}
\end{equation}
For a fixed pump $\mathcal{E}$, we may assume the system evolution to be a functional of the probe shape: $\hat{\rho} = \hat{\rho}[f]$, and we may
expand $\hat{\rho}$ in a Taylor series (in the functional sense) around $f=0$:
\begin{equation}
\hat{\rho}[f] = \sum_{n=0}^{\infty} \hat{\rho}_n[f]\,,
\end{equation}
where $\hat{\rho}_0$ is the unperturbed system evolution (i.e. the
evolution of the system in the presence of the pump only: $f=0$), and
$\hat{\rho}_n$ is $n$-th order in the perturbation:
$\hat{\rho}[\lambda f] =\lambda^n\hat{\rho}[f]$.  The system response
to this perturbation is measured in terms of the expectation value of
an observable $\hat{A}$, which can likewise be expanded:
\begin{equation}
A(t) = {\rm Tr}\left[\hat{\rho}(t)\hat{A}\right] = \sum_{n=0}^{\infty}A_n(t)\,,
\end{equation}
where
\begin{equation}
A_n(t) = {\rm Tr}\left[\hat{\rho}_n(t)\hat{A}\right]\,.
\end{equation}
For sufficiently weak probes, we are only interested in the first term:
\begin{equation}
  \delta A(t) = A(t) - A_0(t) \approx A_1(t) = {\rm Tr}\left[\hat{\rho}_1(t)\hat{A}\right]\, ,
\end{equation}
which is linearly related to $f$ through a pump-dependent response function:
\begin{equation}
A_1(t) = \int_{-\infty}^{\infty}\!\!{\rm d}t'\;f(t')\chi_{\hat{A},\hat{V}}[\mathcal{E}](t,t')\,.
\end{equation}
The response function is given by:\cite{castro2011}
\begin{equation}
\chi_{\hat{A},\hat{V}} [\mathcal{E}](t,t') = i \theta(t-t') \; 
{\rm Tr} \left\lbrace \hat{\rho}_{eq} \left[ \hat{A}_H [\mathcal{E}](t), 
\hat{V}_H [\mathcal{E}](t') \right] \right\rbrace
\label{eq:non-equilibrium}
\end{equation}
Inside the commutator, the operators appear in the Heisenberg representation:
\begin{equation}
  \hat{O}_H[\mathcal{E}] (t) = \hat{U}[\mathcal{E}](t_0,t) \; \hat{O} \; \hat{U}[\mathcal{E}](t,t_0)\,,
\label{eq:heisenberg-picture}
\end{equation}
where $\hat{U}[\mathcal{E}]$ is the time propagation operator in the
presence of the pump only -- hence the functional dependence on
$\mathcal{E}$. We keep this functional dependence explicit in the
notation for $\chi_{\hat{A},\hat{V}} [\mathcal{E}](t,t')$, to stress that
it is a property of both the system (defined by the static Hamiltonian
$\hat{\mathcal{H}}$) \emph{and} of the pump shape $\mathcal{E}$, as
opposed to the conventional response functions, which are only system
dependent. Note also its dependence on two times $t$ and $t'$, which
cannot be reduced to only one by making use of time-translational
invariance, as it is customary when working at equilibrium.


The response itself, $\delta A(t)$, will be a functional of both pump and probe pulses. If we take
its Fourier transform we may write it as:
\begin{equation}
\delta A [\mathcal{E},f](\omega) = \int_{-\infty}^{\infty} dt' f(t')
\chi_{\hat{A},\hat{V}} [\mathcal{E}](\omega,t') \,.
\label{eq:response-w}
\end{equation}
In order to compute the response function, one can use as a probe a delta perturbation,
i.e. $f(t') = \lambda \delta(t'-\tau)$, which permits to identify:
\begin{equation}
\chi_{\hat{A},\hat{V}} [\mathcal{E}](\omega,\tau) = \frac{1}{\lambda}
\delta A [\mathcal{E},\lambda\delta_\tau](\omega)\,.
\end{equation}
The action of such a delta-perturbation applied at the instant $\tau$ 
on the system is given by:
\begin{equation}
\vert\Phi(t\to\tau^+)\rangle = e^{-{\rm i}\lambda \hat{V}} \vert \Phi(\tau)\rangle\,.
\label{eq:delta-application}
\end{equation}
From now on we will restrict the discussion to pure states, since
ensembles are not needed for the results that will be shown below. 
However it can easily be extended to general ensembles.  
We also restrict the discussion to a specific response
function: the dipole-dipole polarizability $\alpha[\mathcal{E}](t,t') =
\chi_{\hat{D},\hat{D}}[\mathcal{E}](t,t')$, where both $\hat{A}$ and
$\hat{V}$ are the atomic or molecular dipole operator $\hat{D}$-- taking into
account that, for the frequencies that we are dealing with, the dipole
of interest is that of the electrons, and the clamped nuclei
approximation can be used. Moreover, we choose to work with light
polarized in the $x$ direction, so that:
\begin{equation}
\hat{D} = -\sum_{i=1}^N\hat{x}_i\,,
\end{equation}
where $N$ is the number of electrons.
The expectation value of this electronic dipole is an explicit functional
of the time-dependent density, and so is its variation:
\begin{equation}
\delta D[\mathcal{E},f](\omega) = -\int\!\!{\rm d}^3r\;\delta n(\vec{r},\omega)x\,,
\end{equation}
where $\delta n(\vec{r},\omega)$ is the Fourier-transformed difference
between the electronic densities obtained with and without the probe
pulse.  This straightforward formula in terms of the density is what
makes TDDFT specially suited for these computations: we may safely
utilize the Kohn-Sham system of non-interacting electrons. The
delta perturbation, Eq.~(\ref{eq:delta-application}), must be
applied to each one of the Kohn-Sham orbitals, and takes the following
form:
\begin{equation}
\varphi(\vec{r},t\to\tau^+) = e^{{\rm i}\lambda x} \varphi_i(\vec{r},\tau)\,.
\end{equation}

The absorption of a particular probe pulse $f$, is determined by the
induced polarization, given by $\delta D[\mathcal{E},f](\omega)$. We will
compute the dynamical polarizability $\alpha[\mathcal{E}](\omega,\tau)$,
which is the polarization induced by a delta perturbation, i.e.:
\begin{equation}
\alpha[\mathcal{E}](\omega,\tau) = \frac{1}{\lambda}\delta D[\mathcal{E},\lambda\delta_\tau](\omega)\,,
\end{equation}
since it
allows to compute any particular response through the integration of
Eq.~(\ref{eq:response-w}), as long as the probe is weak. In
particular, we will look at the imaginary part of
$\alpha[\mathcal{E}](\omega,\tau)$, which is the part responsible for
absorption.
Note, finally, that in a 3D situation the polarizability is
not a scalar but a tensor, since there are three possible light polarization
directions, and three components for the system dipole moment. In most cases,
one is interested in the trace of this tensor, an averaged quantity that corresponds to the
absorption of a randomly oriented sample of molecules.

\subsection{Time-resolved photoelectron spectroscopy}
\label{subsection:pes}

The photoelectron spectra presented in this work are produced within TDDFT using 
the recently introduced Mask Method.~\cite{UDeGiovannini:2012hy}
This method is based on a geometrical partitioning
and a mixed real- and momentum-space time evolution scheme.~\cite{Chelkowski:1998ec} 
In the following, we summarize the main traits of the technique 
(we refer the reader to Ref.~\onlinecite{UDeGiovannini:2012hy}
for a complete description), and demonstrate how
it can be straightforwardly applied to the non-equilibrium situation required
by pump-probe experiments.

In photoemission processes a light source focused on a sample transfers energy 
to the system. Depending on the light intensity electrons can absorb one or 
more photons and escape from the sample due to the photoelectric effect. 
In experiments, electrons are detected and their momentum is measured.
By repeating measurements on similarly prepared samples it is possible to estimate
the probability to measure an electron with a given momentum.
From a computational point of view, the description of such processes for complex systems 
is a challenging problem. The main difficulty arises from the necessity
of describing properly electrons in the continuum.

In typical experimental setups, detectors are situated far away from the sample 
and electrons overcoming the ionization barrier travel a long way before being detected. 
The distances, that electrons travel are usually orders of magnitude larger 
than the typical interaction length scales in the sample.
During their journey towards the detector, and far away from the parent system, they 
practically evolve as free particles driven by an external field.
The solution of the Schr\"odinger equation for free electrons in a time dependent 
external field is known analytically in terms of plane waves as Volkov states. 
It seems therefore a waste of resources to solve the Schr\"odinger equation numerically
in the whole space if a considerable part of the wave function can be described 
analytically.
\begin{figure}
  \centering
  \includegraphics[width=6cm]{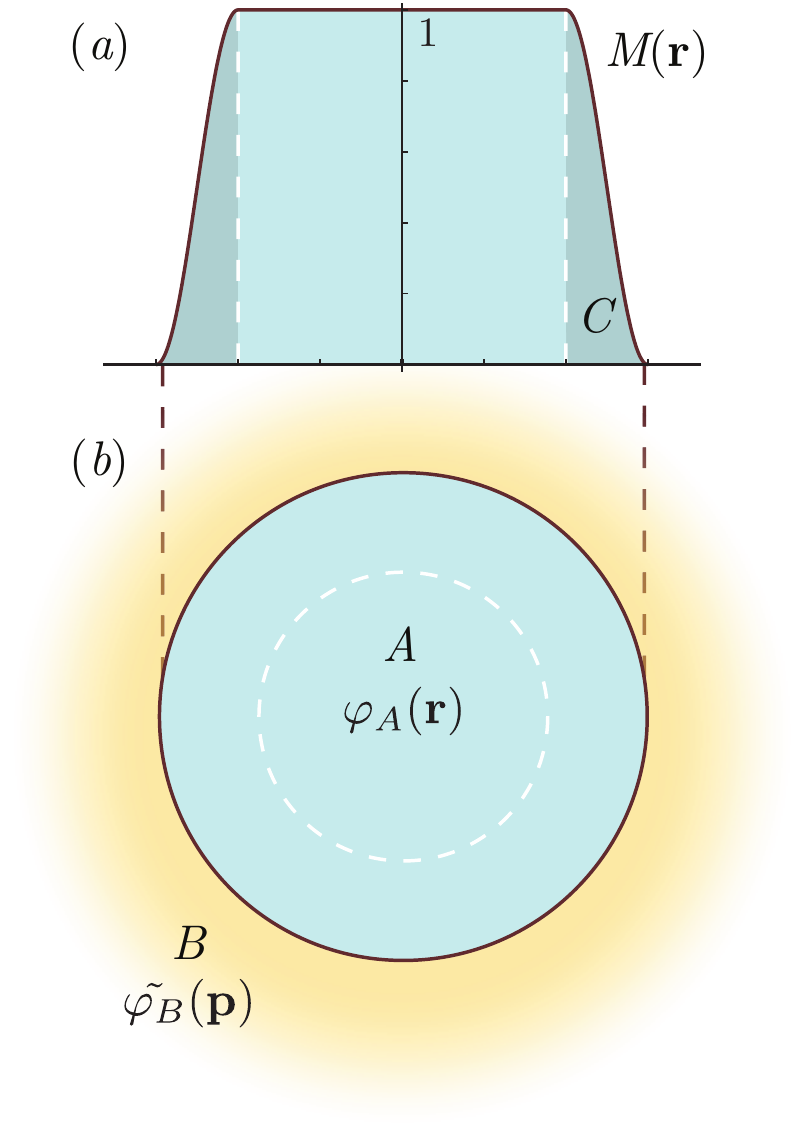}
  \caption{Schematic description of the space partition implemented by the mask method. 
  A mask function (a) is used to implement the spatial partitions (b). 
  In region A (interaction region) the TDKS equations are numerically solved in real space while 
  in B (free propagation region) electrons are evolved analytically as free particles in momentum space.
  Region C is where $\varphi_A$ and $\varphi_B$ overlap. }
  \label{fig:mask_method}
\end{figure}

In order to take advantage of the previous observations 
we partition the space according to the scheme in Fig.~\ref{fig:mask_method} (b).
The space is divided into two regions, $A$ and $B$; the inner region $A$, 
containing the system with enough empty space around, is where electrons are allowed to 
interact with each other and with the system, and region $B$, defined as the complement 
of $A$, is where electrons are non-interacting and freely propagating.
Every KS orbital $\varphi_i({\bf r})$ can be decomposed accordingly with  
$\varphi_i({\bf r})=\varphi_{i}^A({\bf r})+\varphi_{i}^B({\bf r})$, so that $\varphi_{i}^A({\bf r})$ resides mainly
in region $A$ and $\varphi_{i}^B({\bf r})$ mainly in region $B$.

The geometrical partition is implemented by a smooth mask function $M({\bf r})$
defined to be one deep in the interior of $A$ and zero outside (see Fig.~\ref{fig:mask_method} (a)): 
\begin{eqnarray}
\varphi_{i}^A({\bf r}) & = & M({\bf r})\varphi_i({\bf r})\,,
\\
\varphi_{i}^B({\bf r}) & = & (1-M({\bf r}))\varphi_i({\bf r})\,.
\end{eqnarray}
The mask function takes care of the boundary conditions in $A$ by forcing every function to be zero at the border.
In order to give a good description of functions extending over the whole space it is convenient 
to represent $\varphi_{i}^B({\bf r})$ orbitals in momentum space $\tilde{\varphi}_{i}^B({\bf p})$. 

A mixed real and momentum-space time evolution scheme 
can then be easily derived following the geometrical splitting. 
Given a set of orbitals at time $t$ their value at a successive time $t+\Delta t$ is provided by 
\begin{equation}\label{eq:mask_method}  
  \left\{ 
  \begin{array}{ll}
    \varphi_{i}^A(\textbf{r},t+\Delta t) = M({\bf r})e^{- i \hat{H}\Delta t}\varphi_{i}^A(\textbf{r},t)\\ 
    \tilde{\varphi}_{i}^B(\textbf{p},t+\Delta t)= e^{- i \frac{(\mathbf{p}-\mathbf{A}(t))^2}{2} \Delta t}\varphi_{i}^B(\textbf{p},t) + \tilde{\varphi}_{i}^A(\textbf{p},t+\Delta t)
\end{array} \right. 
\end{equation}
with $\hat{H}$ being the effective single-particle TDDFT Hamiltonian, $\mathbf{A}(t)$ the total external time dependent 
vector potential (the coupling with the external field is conveniently expressed in the velocity gauge), and
\begin{equation}
  \tilde{\varphi}_{i}^A(\mathbf{p},t+\Delta t)=\frac{1}{(2\pi)^{3/2}}\int d\mathbf{r} 
    (1-M({\bf r}))e^{- i \hat{H}\Delta t}\varphi_{i}^A(\textbf{r},t) e^{i \mathbf{p}\cdot \mathbf{r}} \,,
\end{equation}
constituting the portion of electrons leaving the system at time $t+\Delta t$.
At each iteration in the evolution the outgoing components of $\varphi_{i}^A(\textbf{r})$ are suppressed in the 
interaction region by the multiplication with $M(\mathbf{r})$ while being collected as plane waves in 
$\tilde{\varphi}_{i}^B(\textbf{p})$ via $\tilde{\varphi}_{i}^A(\textbf{p})$. The resulting momentum space wavefunctions are then 
evolved analytically simply by a phase multiplication. 

The advantage of using such an approach resides in the fact that we can conveniently
store the wavefunctions on a spatial grid inside $A$ while treating wavefunctions in 
$B$ (and therefore the tails extending to infinity) as free-electrons in momentum space. 
Moreover the mask function introduces a region $C$, where
the wavefunctions in $A$ and $B$ overlap (see Fig.~\ref{fig:mask_method}) and that acts as matching layer.
In spite of the fact that, from a theoretical point of view, the matching between inner and outer region could be performed
on a single surface, from a numerical point of view, having a whole region to perform the matching is more 
stable and less influenced by different choices of spatial grids.

From the momentum components of the orbitals in $B$ it is possible to 
evaluate the momentum-resolved photoelectron probability distribution
as a sum over the occupied orbitals
\begin{equation}
	P(\mathbf{p}) \approx \lim_{t \rightarrow \infty}\sum_{i=1}^{occ} | \tilde{\varphi}_i^{B}(\mathbf{p},t) |^2\,,
	\label{eq:p-resolved}
\end{equation}
the limit $t \rightarrow \infty$ ensuring that all the ionized components are collected.
This scheme is entirely non-perturbative; in a pump-probe setup, it does not assume
linearity in either pump or probe. Therefore, it can be applied in the same manner
when two pulses are present than with one pulse only, as it was
shown in Ref.~\onlinecite{UDeGiovannini:2012hy}.
Like in Sec.~\ref{subsection:tas} we can generalize the previous derivation to address transient 
photoelectron spectroscopy (spin-, angle- and energy-resolved) in practice by 
employing a pump-probe scheme and performing numerical simulations
with two time delayed external pulses. A TRPES map is then generated
by performing a computation for each different time delay.
 
From $P(\mathbf{p})$ several relevant quantities can be calculated. The energy-resolved photoelectron 
probability $P(E)$, usually referred to as photoelectron spectrum (PES), can be obtained by integrating $P(\mathbf{p})$
over solid angles
\begin{equation}
	P(E=\mathbf{p}^2/2) = \int_{4\pi} d\Omega_p\, P(\mathbf{p})\, .
	\label{eq:PES_def}
\end{equation}
The angular- and energy-resolved photoelectron probability $P(\theta,\phi,E)$, or photoelectron angular distribution (PAD),
can easily be evaluated by expressing $P(\mathbf{p})$ in polar coordinates with respect to a given azimuth axis.    

It is noteworthy that during the evolution defined in Eq.~\ref{eq:mask_method} the part of the 
density contained in $A$ transferred to $B$ is not allowed to return. 
Clearly, in cases where the external field is strong enough to produce electron orbits 
crossing the boundary of $A$ and backscattering to the core the mask method provides a poor approximation. 
In these cases a bigger region $A$ or a more refined scheme must be employed~\cite{UDeGiovannini:2012hy}.
The laser fields employed in this work are weak enough that   
we can safely assume region $A$ to always be sufficiently large to 
contain all the relevant electron trajectories.

\section{Results}

The theory described in the previous section has been implemented in
the {\tt octopus} code~\cite{octopusnote}; we refer the reader to
Refs.~\onlinecite{Marques200360,PSSB:PSSB200642067} for the essential points
of the numerical methodology.

In order to simplify the illustration of the results a clamped ion
approximation has been used in the calculations for the molecular case.
Further inclusion of the ionic motion could be done at the semi-classical level 
with Ehrenfest dynamics~\cite{Marques200360,Alonso:2008fz} 
-- already implemented in the code -- and without any essential modification 
to the theory presented.

\subsection{One-dimensional model Helium}
\label{sec:he_1d}

As first example we study the absorption spectrum of an excited one-dimensional 
soft-Coulomb Helium atom. This is an exactly solvable model that provides a useful 
benchmark to test different approximations. 
We will first discuss the exact solution, and later apply TDDFT.
A more realistic 3D model will be presented in the next section.

The 1D model of the Helium atom is defined by the following Hamiltonian:
\begin{equation}
H(t)= T + V_{ext}(t)+V_{ee}\,,
\end{equation}
where
\begin{equation}
V_{ext}= -\frac{2}{\sqrt{1+x_1^2}} - \frac{2}{\sqrt{1+x_2^2}} + \mathcal{E}(t)(x_1+x_2)
\end{equation}
is the external potential: the electron Coulomb interaction $1/|x|$ is softened
to $1/\sqrt{1+x^2}$. 
The coupling with the external time-dependent field $\mathcal{E}(t)$
is expressed in length gauge, and electrons are confined to move along
the $x$ direction only. Finally, 
\begin{equation}
T=-\frac{1}{2} \left( \frac{\partial^2}{\partial x_1^2} + \frac{\partial^2}{\partial x_2^2} \right)
\end{equation}
is the kinetic energy, and the electron-electron interaction is
\begin{equation}
V_{ee} = \frac{1}{\sqrt{1+(x_1-x_2)^2}}\,.
\end{equation}
This model is numerically solvable given the exact mapping 
discussed in Refs.~\onlinecite{Helbig:2011kk,Helbig:2011bxa}, where it is proved that 
the many-body problem of $N$ electrons in one dimension is equivalent to that of one 
electron in $N$ dimensions. 
The wave functions and other necessary functions are
represented on a real space regular grid; a squared (linear for
one-dimensional TDDFT) box of size $L=200$ a.u.  and spacing $\Delta
x=0.2$ a.u. has been employed in all the calculations.

\begin{figure}
    \centering
    \includegraphics[width=\columnwidth]{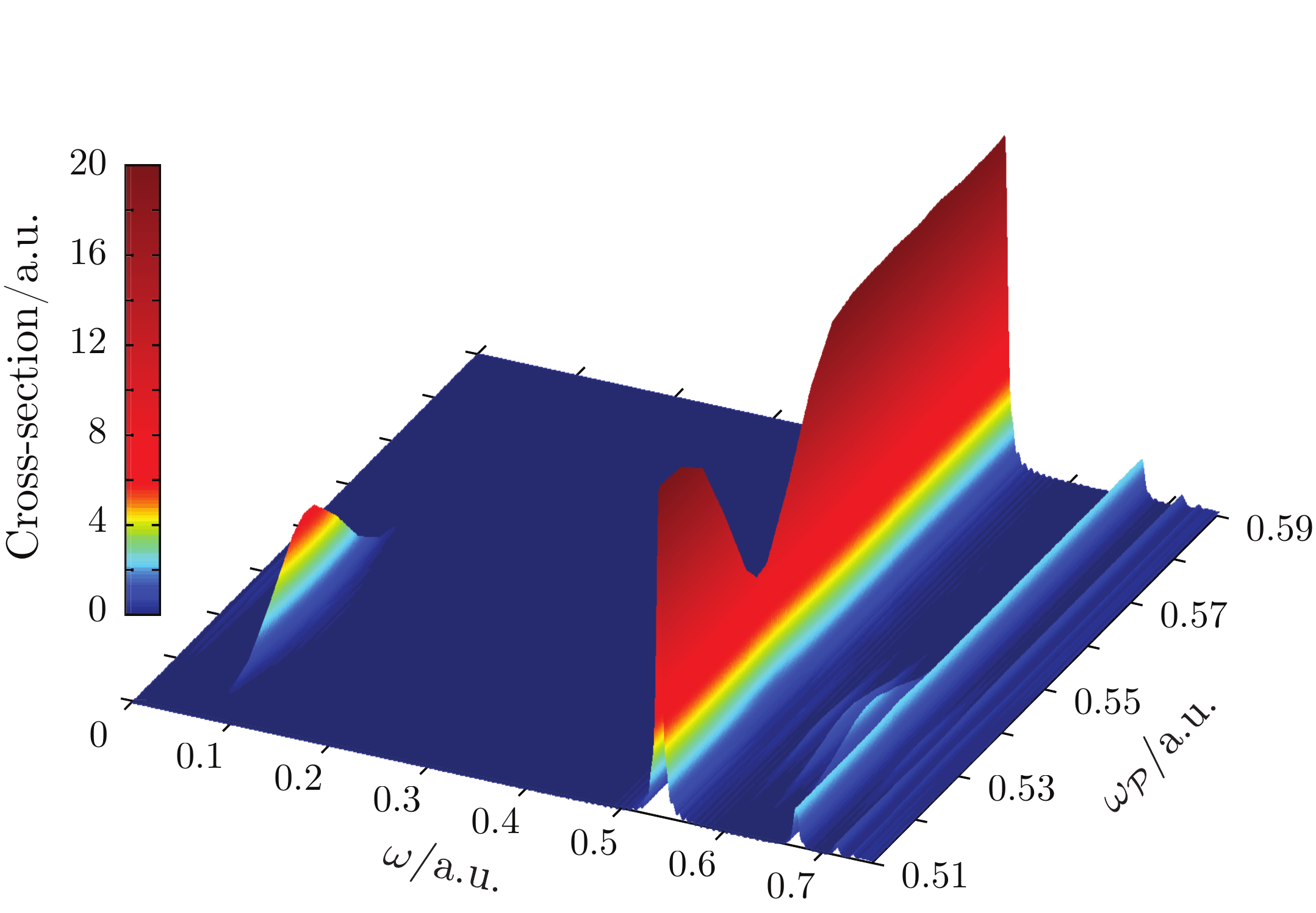}
    \caption{Out of equilibrium absorption spectrum as function of the pump laser frequency for 
    one-dimensional Helium. The system is driven out of equilibrium by 45 cycle $\sin^2$ envelope laser pulses of 
    intensity $I=5.26\times 10^{11}$~W/cm$^2$, at different carrier frequencies and then probed right after. 
    Maximal response is observed for frequencies close to the first optical transition $\omega_{\mathcal{P}}=0.533$ a.u..}
    \label{fig:He_1D-2D_map}
\end{figure}
In order to illustrate how an external field can modify the optical properties of a system 
in Fig.~\ref{fig:He_1D-2D_map} we show a scan of the non-equilibrium absorption spectrum 
generated by a 45 cycle $\sin^2$ envelope pulse with intensity $I=5.26\times 10^{11}$ W/cm$^2$ 
at different angular frequencies 0.51~[a.u.]~$\leq \omega_{\mathcal{P}} \leq$~0.59~[a.u.].
The plot displays ${\rm Im}[\alpha[\mathcal{E},\tau](\omega,\omega_{\mathcal{P}})]$, choosing $\tau$ right at the end of
the pump pulse.

It can be seen how the absorption around the first excitation
frequency 0.533 a.u. is strongly diminished when the frequency of the pump
resonates with that frequency. In that situation,
an absorption peak also appears around the excitation frequency corresponding to the
transition from the first to the second excited state, in our case at 0.076 a.u.
This behavior is a direct consequence of the fact that the laser is pumping the system to the 
first excited state and this process is more efficient for a field tuned with the excitation 
energy. The absorption spectrum is therefore a mixture of the one corresponding to the 
unperturbed ground state and that of the first excited state.

\begin{figure}
  \centering
  \includegraphics[width=\columnwidth]{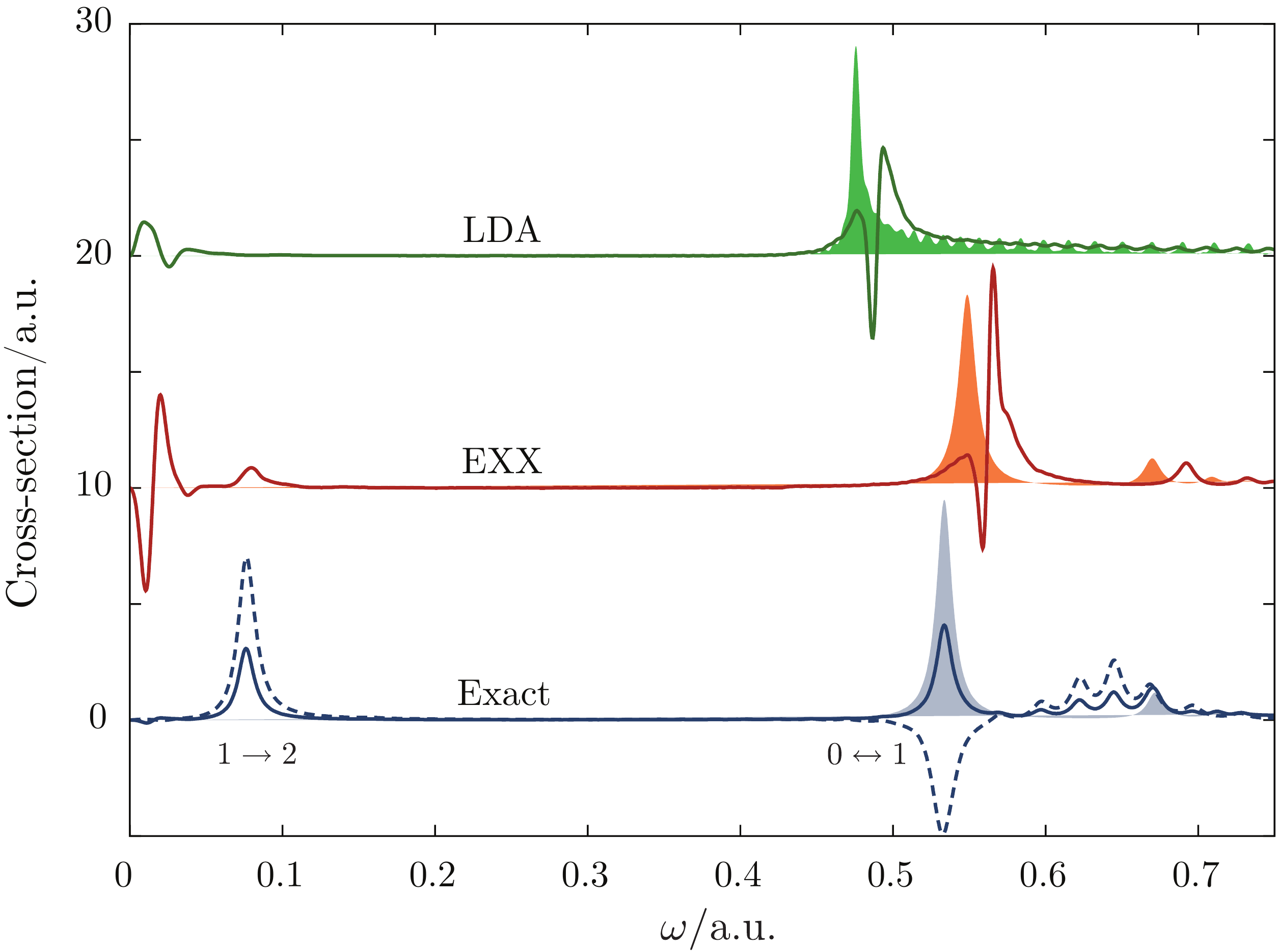}
    \caption{Comparison of absorption spectra calculated in different approximations for a one-dimensional Helium model. 
    The filled curves are the spectra for the unperturbed systems while the solid lines are the spectra  
    of the system excited by a laser as in Fig.~\ref{fig:He_1D-2D_map} resonant with the 
    first allowed optical transition: exact time-dependent Schr\"odinger equation $\omega_{\mathcal{P}}=0.533$ a.u. (in blue), 
    EXX $\omega_{\mathcal{P}}=0.549$ a.u. (in red), and LDA 
    $\omega_{\mathcal{P}} =0.475$ a.u. (in green). The dashed blue line is the absorption of the system perturbed by a 180 cycle laser
    and probed at $t=30.62$~fs, where the population on the excited state is maximal. 
    The lines have been shifted by a vertical constant to facilitate the comparison between results.}
  \label{fig:he1d-spectrum}
\end{figure}
In order to analyze this point further, a cut at the resonant frequency is displayed in the lower (blue) curves
of Fig.~\ref{fig:he1d-spectrum}. The filled curve represents the spectrum obtained from the system in its 
ground state while the solid line corresponds to the spectra of the system excited by a laser pulse with a 
frequency resonant with the first optical transition and probed after the perturbation.
By direct comparison of the two spectra it is easy to discriminate the peaks associated with the ground-state absorption from 
the ones characterizing the absorption from the excited state. 
In particular, the peaks related to the ground-state absorption 
are located at the energies $\omega_{0\rightarrow 1}=\epsilon_1-\epsilon_0=0.533$ a.u. 
corresponding to the transition from the ground ($\epsilon_0=-2.238$ a.u.) to the first excited state  
($\epsilon_1=-1.705$) and $\omega_{0\rightarrow 3}=\epsilon_3-\epsilon_0=0.672$ a.u. corresponding to the third 
excited state ($\epsilon_3=-1.566$ a.u.) -- the direct excitation of the second excited state is forbidden by symmetry.
The solid curves show
fingerprints of the population of the first excited state, namely the peaks corresponding
to transitions from that first excited state to others: 
in particular, the peak appearing at the low energy 
$\omega_{1\rightarrow2}=\epsilon_2-\epsilon_1=0.076 $ a.u. is associated with the transition from the first excited
state $\epsilon_1$ to the second one $\epsilon_2=-1.629$ a.u.

These spectra contain information that is not contained in the
equilibrium ones.  For example, let us consider the
spectra that would be produced by each single eigenstate, given by the
state-dependent dynamical polarizabilities, which may be
written in the sum-over-states form as:
\begin{eqnarray}
  \label{eq:sos}
  \alpha^{(i)}(\omega)=\sum_{j\neq i} & &\left[  \frac{\vert\langle\Psi_i|\hat{D}|\Psi_j\rangle\vert^2}
  {\omega - (\epsilon_j-\epsilon_i)+i0^+ } \right. \\
  &-&\left. \frac{\vert\langle\Psi_i|\hat{D}|\Psi_j\rangle\vert^2}
  {\omega + (\epsilon_j-\epsilon_i)+i0^+ } \right]\nonumber\,.
\end{eqnarray}
The poles of this function provide us with the eigenvalue differences $\epsilon_j-\epsilon_i$;
if this value is positive, the corresponding term is associated with a photon-absorption process;
if it is negative, with a stimulated emission term. The weight associated with each one of these
poles provide us with the dipole coupling matrix elements $\langle\Psi_i|\hat{D}|\Psi_j\rangle$.

During the time evolution, the wavefunction can be expanded on the basis of
eigenstates of the unperturbed system $\Psi(t)= \sum_i\eta_i(t)\Psi_i$.
When the system is probed at a certain time $t$, the resulting spectrum
can be thought as a linear combination of the spectra produced by each single eigenstate.
An analysis of the transient spectrum may therefore provide information about the mixing
weights $\eta_i$, and about excitation energies and dipole couplings between excited states -- information
that is absent in equilibrium ground-state linear response.

In our case, we find by direct projection of the time-dependent wave function onto the eigenstates, that the system after the pulse is composed mainly of the
ground and the first excited state with weights, $|\eta_0|^2=|\langle \Psi_0|\Psi(t)\rangle|^2=0.7120$  
and $|\eta_1|^2=|\langle \Psi_1|\Psi(t)\rangle|^2=0.2876$.
The same information can be recovered by comparing the perturbed and the unperturbed spectrum at $E_{0\rightarrow 1}$.
At this energy we only have the contribution coming from $\Psi_0\rightarrow \Psi_1$ and 
its inverse $\Psi_1\rightarrow \Psi_0$. The peak height of the perturbed spectrum after the laser pulse $h_t$ 
is therefore a combination of the heights associated with the ground 
$h_0$ and the excited $h_1$ states: $h_t=|\eta_0|^2 h_0+|\eta_1|^2 h_1$.
At this energy $h_0=-h_1$ due to the nature of the transition $1\rightarrow 0$ 
the ratio $\alpha=h_t/h_0=|\eta_0|^2-|\eta_1|^2=0.4258$ thus gives direct information on the 
difference of the mixing weights.
Complementing this information with a two-level system assumption $|\eta_0|^2+|\eta_1|^2=1$ we obtain 
$|\eta_0|^2=(1+\alpha)/2=0.7129$ and $|\eta_1|^2=(1-\alpha)/2=0.2871$ in good agreement with the results 
calculated by direct projection of the wavefunction. 

\begin{figure}
  \centering
  \includegraphics[width=8cm]{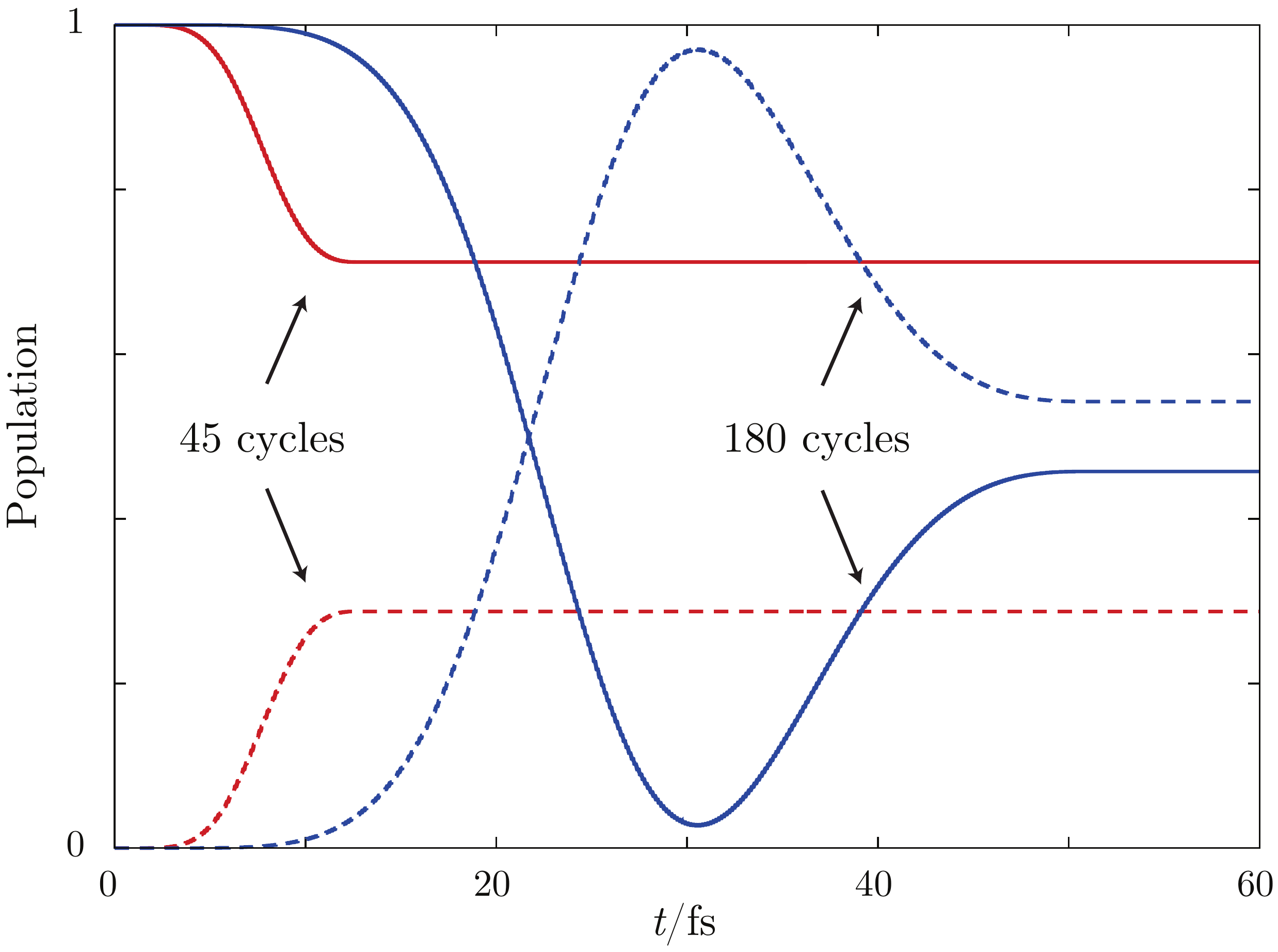}
    \caption{Exact population on the ground $|\eta_0(t)|^2=|\langle \Psi_0|\Psi(t)\rangle|^2$ 
    (solid lines) and first excited $|\eta_1(t)|^2=|\langle \Psi_1|\Psi(t)\rangle|^2$ 
    (dashed lines) states as a function of time for different laser pulses. In red a 45 cycles pulse 
    with parameters as in Fig.~\ref{fig:he1d-spectrum}, and in blue a longer 180 cycles pulse
    with the same parameters.}
  \label{fig:he1d-populations}
\end{figure}
In Fig.~\ref{fig:he1d-populations} we display the population weights for two different laser pulses.
The red lines correspond to the same laser pulse as in Fig.~\ref{fig:he1d-spectrum} while the blue 
lines pertain to a four times longer laser with the same parameters (intensity, envelope shape and carrier frequency)
and 180 optical cycles.
For both lasers the populations of both the ground and first excited state at each time almost sum to one,
indicating an 
essential two-level dynamics. 
In the case of the long pulse we observe a maximum (minimum) of the population over the 
excited (ground) state at $t=30.62$~fs. This behavior can be understood in terms
of Rabi oscillations. 

A Rabi oscillation is a fluctuation behavior of 
states occupation occurring  due to the interaction of an oscillatory 
optical field in resonance with a two-level system.~\cite{Allen:1975} 
The occupation probability alternates with the Rabi frequency 
$\Omega(t)=f(t) \mu_{0\rightarrow 1}$, where $\mu_{0\rightarrow 1}$ is the 
dipole transition matrix element between the states and $f(t)$ is the electric field envelope.
Extremal points of the populations should be located at times where the 
pulse area $\Theta(t)=\int_{-\infty}^{t}d\tau\, f(t) \mu_{0\rightarrow 1}$ is an 
integer multiple of $\pi$, $\Theta(t)=n\pi$.
With the numerically calculated matrix element $\mu_{0\rightarrow 1}=1.11$ a.u. the 
first maximal population of the excited state is expected at $t=30.65$~fs, in good 
agreement with what is observed. The absorption spectrum at this time, as shown in Fig~\ref{fig:he1d-spectrum}
(dashed blue line), displays a considerable enhancement at $\omega_{1\rightarrow 2}$ and a negative 
emission peak at $\omega_{0\rightarrow 1}$ as expected from a pure excited state.

It is interesting to study the same model with TDDFT instead of with an exact treatment
in order to address the performance of available (mainly static) xc-functionals.
In Fig.~\ref{fig:he1d-spectrum} we display results obtained with TDDFT, 
employing two different exchange-correlation (xc) functional approximations:
exact exchange~\cite{Kummel:2003hpa} (EXX) in red, 
and one-dimensional local density approximation~\cite{Casula:2006fi} (LDA) in green.
The calculations  were  performed in the adiabatic approximation using the same parameters as in the exact case. 
The laser frequency was tuned to match the first optical transition appearing at: $\omega_{\mathcal{P}}=0.549$ a.u. for EXX, and
$\omega_{\mathcal{P}}=0.475$ a.u. for LDA. 

The unperturbed spectrum (solid curve) provided by EXX is in good agreement with the exact calculation, and 
the perturbed one qualitatively reproduces the exact result. In particular the new
peak appearing at low energy associated with the transition $1\rightarrow 2$ is well represented. 
In contrast LDA is only capable of reproducing one peak both for the perturbed and unperturbed
cases. 
This is due to the known problem of asymptotic exponential
decay of the functional that in this one-dimensional example supports only a single bound
excited state.

A common feature of both approximations is constituted by the presence of negative values in the perturbed
spectra. This can be tracked down to the lack of memory in the adiabatic xc-functional approximation~\cite{Fuks:2011hp}.
The lack or wrong memory dependence in the functional results in slightly displaced absorption and emission peaks
associated with the same transition.
This fact, analyzed in the light of Eq.~\ref{eq:sos}, results, at the transition energy, in a sum of two Lorentzian curves with different sign and slightly 
different centers. This explains why we get two inverted peaks 
where we should have only a single one going from positive to negative strength as we populate the 
excited state -- as shown by the exact (blue) curves in Fig.~\ref{fig:he1d-spectrum}.

\subsection{Helium atom in 3D}
\label{sec:helium3d}

In this section we study the real Helium atom.
We employed the EXX functional and discretized TDDFT equations 
on a spherical box of radius $R=14$ a.u., spacing $\Delta r=0.4$ a.u. 
and absorbing boundaries 2 a.u. wide. 

We begin by investigating the changes in absorption of He 
under the influence of an external UV laser field driving the system with the frequency of the 
first dipole-allowed excitation. 
To this end we used a 45 cycle $\sin^2$ laser pulse in velocity gauge with carrier $\omega_{\mathcal{P}}= 0.79$~a.u. 
resonant with the $1s^2\rightarrow 1s2p$ transition, of intensity $I=2.6\times10^{12}$~W/cm$^2$
polarized along the $x$-axis.
\begin{figure}
	\includegraphics[width=8cm]{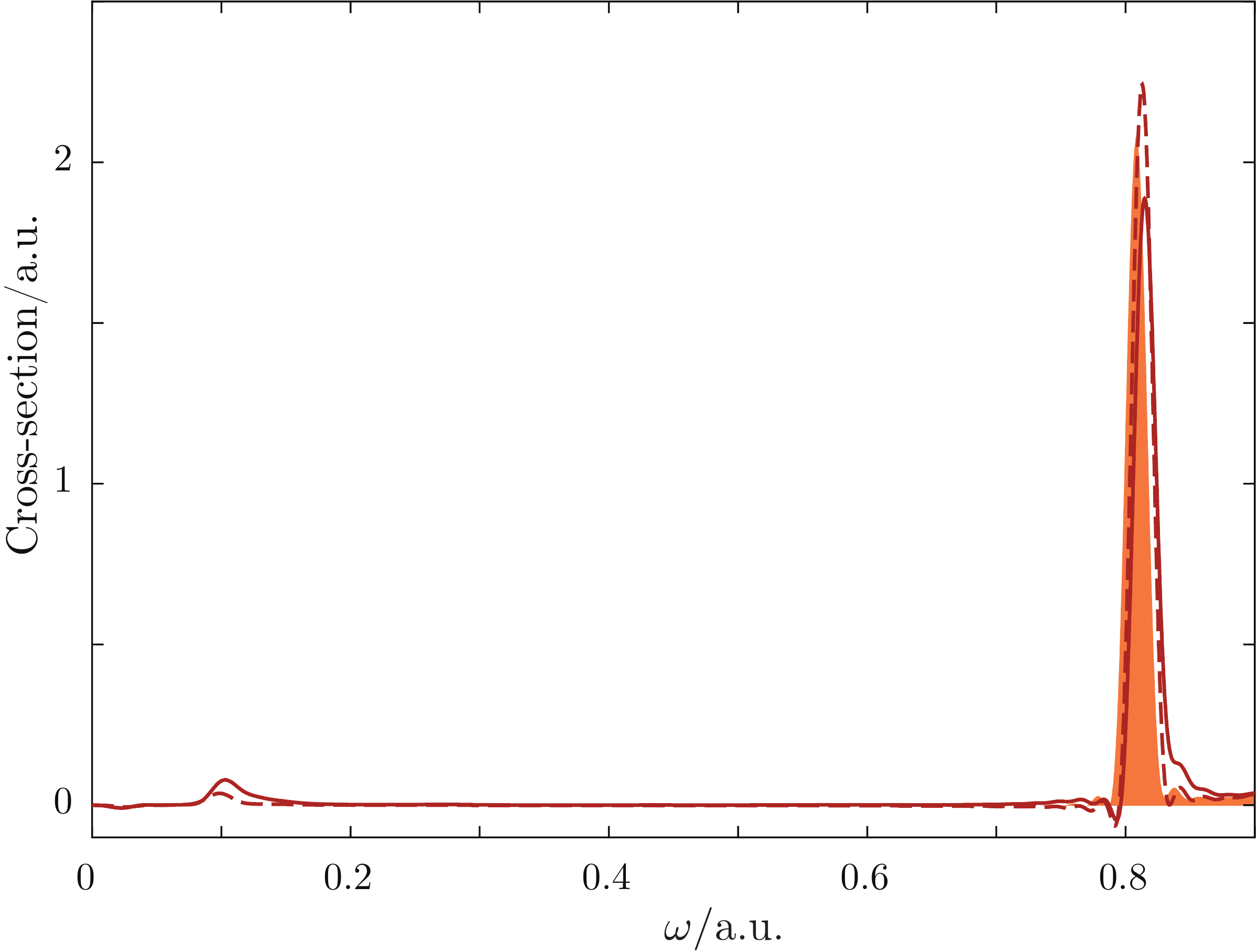}
	\caption{Comparison of the absorption spectra of unperturbed (filled curve) and perturbed He atom 
  probed at $\tau= 5.39$ (solid line) and after the end of the pulse $\tau= 8.68$~fs (dashed line).
  The spectrum range is below the ionization threshold.
  The atom is excited by a 45 cycle $\sin^2$ envelope laser pulse polarized along the $x$-axis 
  with carrier $\omega_{\mathcal{P}}= 0.79$ a.u.
  resonant with the first optical transition, intensity $I=2.6\times10^{12}$ W/cm$^2$.}
	\label{fig:he-abs-pp-cut}
\end{figure}
In Fig.~\ref{fig:he-abs-pp-cut} we show a comparison of the absorption spectrum for the 
unperturbed atom (filled curve) and the perturbed one probed with a delta perturbation right after 
the pump pulse at $\tau=8.68$~fs (dashed line).
The comparison presents many traits similar to the ones
discussed in Sec.~\ref{sec:he_1d} for the one-dimensional Helium model.
In particular, fingerprints of the population of the first excited state
can be observed in the appearance of a peak in the gap at $\omega_{2p\rightarrow 3s}=0.079$~a.u. associated 
with the transition $1s2p\rightarrow 1s3s$.  
The second peak, associated with the transition $1s^2\rightarrow 1s2p$, 
presents height changes correlated with the former one. We also obtain the small artifacts, 
such as the energy shifts and the negative values attributed to the xc-kernel memory dependence discussed previously.

\begin{figure}
	\includegraphics[width=\columnwidth]{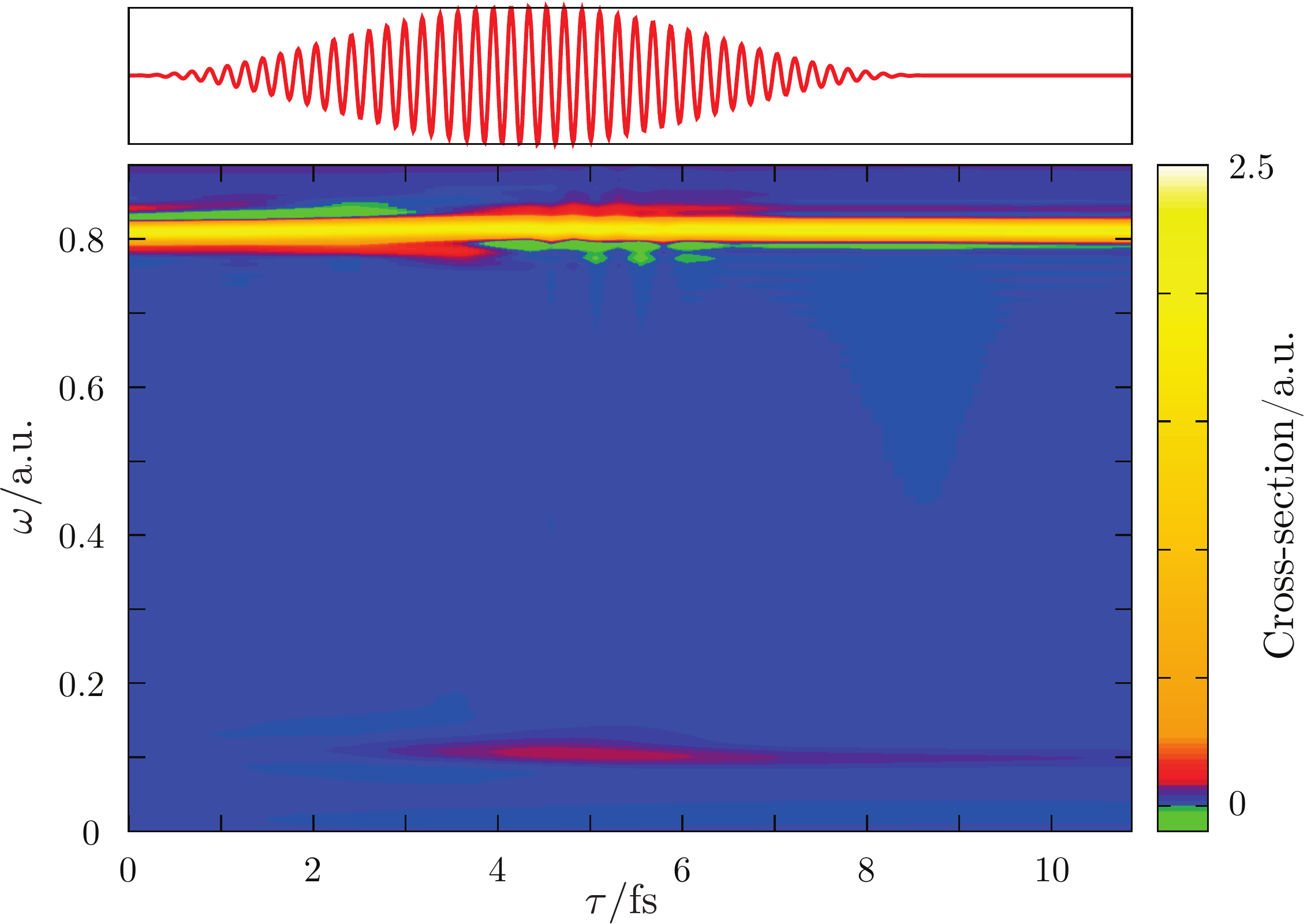}
	\caption{Helium transient absorption spectrum scan for different time delays. Pump laser pictured in 
  the upper panel is the same as in Fig.~\ref{fig:he-abs-pp-cut}.}
	\label{fig:map-delay_time}  
\end{figure}
Additional details on the excitation process can be acquired by expanding 
the time dimension of the absorption spectrum.
In Fig.~\ref{fig:map-delay_time} the time resolved absorption spectrum (TAS) map is displayed.
The map  was produced by probing the system at different time delays. 
As the delay is increased we observe the build-up of the peak associated with the 
state being pumped by the laser pulse at $\omega_{2p\rightarrow 3s}$. 
This changes are reflected in the oscillations of the ground state first optical peak at $\omega_{1s\rightarrow 2p}$.  
In TDDFT the knowledge of the wavefunction is lost in favor of the density,
which does not allow us to do a population analysis based on simple wave function projection
The transient absorption spectrum, on the other hand, is an explicit density functional, and
its computation with TDDFT may help us to understand the evolution of the state populations.

The peak appearing in the gap presents a maximum at $\tau=5.39$~fs that emerges before the end 
of the pump pulse ($\tau=8.68$~fs). 
This peak is associated only with the transition from the $1s2p\rightarrow 1s3s$  and therefore its 
height is proportional to the $2p$ excited state population.
The oscillation can then be interpreted in terms of Rabi physics as discussed in Sec.~\ref{sec:he_1d}.  

\begin{figure}
    \includegraphics[width=\columnwidth]{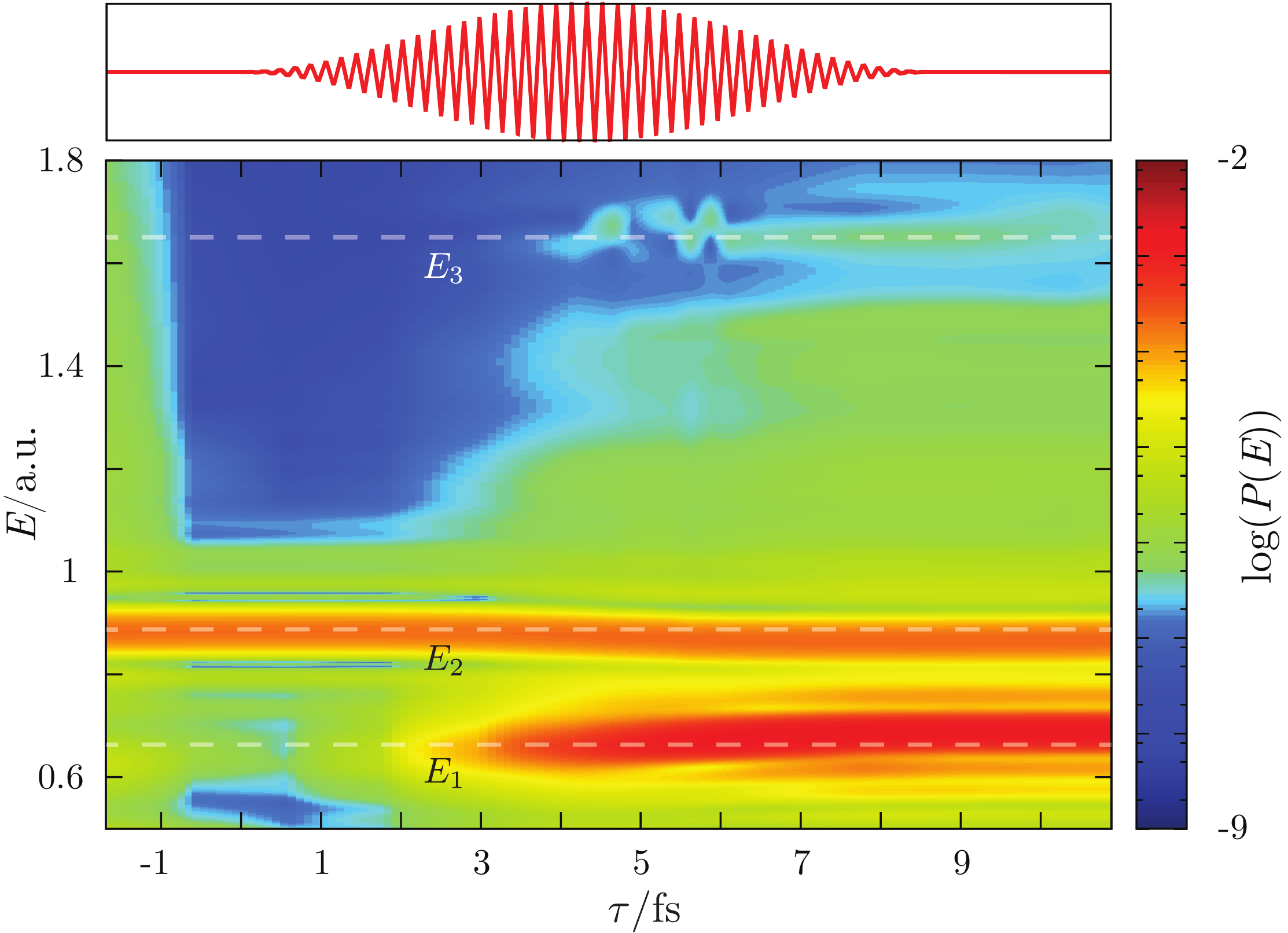}
    \caption{Helium transient photoelectron spectrum in logarithmic scale. The pump laser (upper panel) is
    the same as in Fig.~\ref{fig:he-abs-pp-cut} and the probe is a 40 cycles trapezoidal laser pulse with 8 cycles ramp,  
    $\omega_p=1.8$ a.u., $I=5.4\times10^{9}$ W/cm$^2$ aligned with the pump pulse. }
    \label{fig:He_PES_pp}
\end{figure}
Further insight can be achieved investigating the photoemission properties
of the system. 
In Fig.~\ref{fig:He_PES_pp} we show the TRPES map, as calculated in a pump-probe set up.
Photoelectrons are calculated with the technique outlined in Sec.~\ref{subsection:pes}.
The pump pulse is the same as the one employed for TAS.
The probe is a 40 cycles trapezoidal laser pulse (8 cycles ramp) with carrier frequency $\omega_p=1.8$ a.u.,  
intensity $I=5.4\times10^{9}$ W/cm$^2$, polarized along the $x$-axis and is weak enough to discard non-linear effects.
We performed a scan for different time delays, measuring each delay as the difference from 
the probe center to the beginning of the pump. Negative delays correspond to the situation where 
the probe precedes the pump.
Moreover, in order to include all the relevant trajectories a spherical box of $R=30$ a.u. was 
employed, and photoelectrons were recorded only during the up-time of the probe pulse.

The TRPES map in Fig~\ref{fig:He_PES_pp} shows three main features at $E_1=0.66$~.a.u, $E_2=0.88$~a.u. and $E_3=1.67$~a.u.. 
In our case the probe pulse is weak, and photoelectrons escaping the system undergo 
photoelectric-effect energy conservation. 
A bound electron can absorb a single photon and escape from the atom with 
a maximum kinetic energy $E=\omega_{p}-I_P$, where $\omega_{p}$ is the 
probe carrier frequency and $I_P$ is the field-free ionization energy. 
The ionization potential can be evaluated in DFT as the 
negative energy of the highest occupied KS orbital (HOMO) $I_P=\epsilon_{2s}=0.92$~a.u.
Thus the peak appearing at $E_2$ is energetically compatible with 
photoelectrons emitted from the $2s$ level: $E_2=\omega_p-I_P$.
Consistently, this peak is the only one appearing at negative delays where the 
pulses do not overlap. Moreover, the peak strength is weakly varying with the delay
while slightly shifting towards lower values around 3~fs in accordance with 
TAS findings. At about the same delay time the peak at $E_3$ begins to emerge.
This peak corresponds to emission from the pump-excited 
$2p$ state $E_3=\omega_{\mathcal{P}} + \omega_p -I_P$. It is a process where 
the atom, initially in the ground state, absorbs a photon from the pump
and gets excited to the $2p$ bound state. The subsequent absorption of a probe photon 
frees the electron into the continuum.
The peak at $E_1$ is understood in terms of pump photons only: $E_3=2 \omega_{\mathcal{P}} -I_P$.
The ionization mechanism shares the first step with the $E_3$ process, namely a $2s\rightarrow 2p$
excitation produced by the absorption of a $\omega_{\mathcal{P}}$ photon.  
In the second step the electron is directly excited to a continuum state by the 
absorption of a second $\omega_{\mathcal{P}}$ photon.
In the linear regime, the direct photoionization cross-section decays exponentially with energy
\cite{Fojon:2006ut}. 
For this reason and due to the disparity in intensity between pump and probe 
this ionization channel is by far the most favorable one.  

\begin{figure}
    \includegraphics[width=\columnwidth]{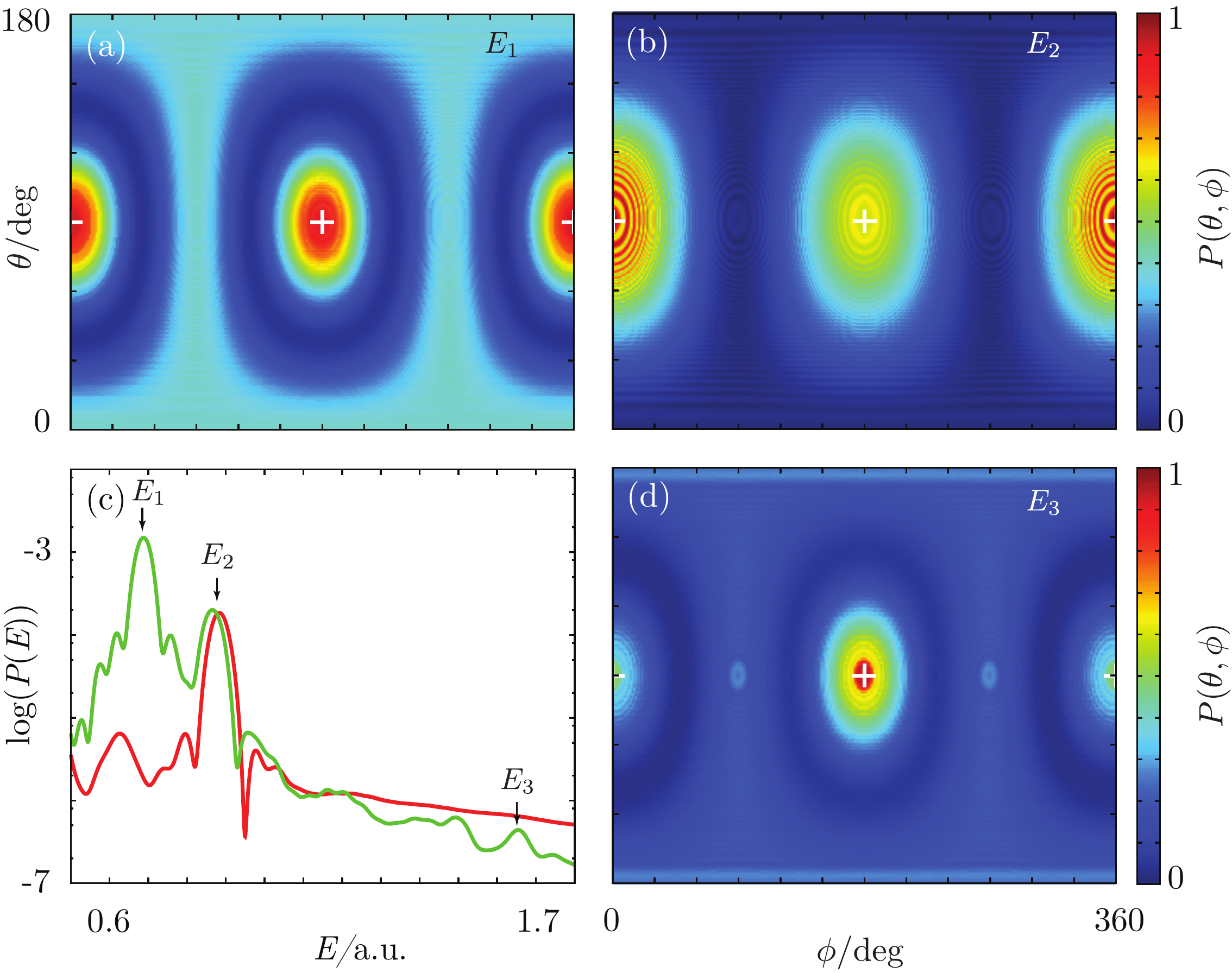}
    \caption{Energy- and angular- resolved photoelectron spectra for Helium 
    at fixed delay $\tau = 8.99$~fs. Panel (c) displays a logarithmic 
    scale PES $P(E)$ comparison at fixed delays $\tau= -1.69$~fs (red) and  $\tau = 8.99$~fs (green). 
    The other panels depict normalized PADs $P(\theta,\phi, E)$ with polar coordinates referred to
    axis $z$ at fixed delay $\tau = 8.99$~fs and energy: (a) $E_1=0.66$~a.u., 
    (b) $E_2=0.88$~a.u., and (d) $E_3=1.67$~a.u..
    White crosses mark the intersection between the probe polarization axis and the
    cutting sphere. }
    \label{fig:He_PES_panel}
\end{figure}
In direct photoemission processes, the photoelectron angular distribution (PAD)
contains information about the electronic configuration of the ionized state.~\cite{Puschnig:2009ho} 
In order to support the energetic arguments PADS $P(\theta,\phi, E)$
at $\tau = 8.99$~fs are presented in Fig~\ref{fig:He_PES_panel} (a), (b), (d) 
together with cuts on TRPES map at $\tau= 0$~fs and $\tau = 8.99$~fs (b).
For each energy marked in Fig.~\ref{fig:He_PES_panel} (c)
we perform spherical cuts of the photoemission probability on energy shells at $E=E_1$, $E_2$, $E_3$.
Each cut is then plotted in polar coordinates with $\theta$ being the angle from the $z$-axis and $\phi$
the angle in the $xy$-plane measured from the $x$-axis. Intersection of the
lasers polarization axis with the sphere are marked with a white cross.

Fig~\ref{fig:He_PES_panel} shows clearly that photoelectrons
at $E_1$ (a) and $E_3$ (d) have similar nature compared to $E_2$ (b), in agreement with the energy analysis.    
Electrons emerging with a kinetic energy of $E_2$ are emitted from 
a $2s$ state, and symmetry of the orbital is imprinted in the photoelectrons angular distribution. 
In order to understand the PAD features it must be taken into account that
2$s$ electrons are perturbed by a laser with a specific polarization direction
that breaks the rotational symmetry. The laser transfers maximal kinetic energy 
along directions parallel with the polarization and minimal along the perpendicular plane
and, if non-liner effects can be discarded, it induces a geometrical factor of the 
form $|\mathbf{A}\cdot \mathbf{p}|$ where $\mathbf{A}$ is the polarization direction
and $\mathbf{p}$ the electron momentum.
For this reason electron emitted along $\phi=90^{\circ}$, and $270^{\circ}$ are strongly suppressed, 
and panel (b) is compatible with the spherical symmetry of a $2s$ state.

In panel (a) electrons are excited to a $p$ state and then ejected into the continuum by the absorption
of two pump photons. The PAD displays marked emission maxima for the direction aligned with 
the laser polarization (indicated by white crosses). 
The extension in $\theta$ is narrower compared with the $2s$ emission in panel (b) 
consistently with ionization from a $p_x$ orbital. 
Of the three degenerate $p$ orbitals the $p_x$ is the one producing the strongest 
response. Signatures of $p_y$ and $p_z$ response can be identified in the non
vanishing PAD on the $yz$-plane around $\phi=90^{\circ},270^{\circ}$.
Such perpendicular response indicates a degree of non-linearity induced by 
the pump.
Similar considerations hold for panel (b) where the $p$ state excited by the
pump is probed with $\omega_p$. As before the emission is mainly  
from a $p_x$ state. 

\subsection{Ethylene molecule} 
\label{sub:ethylene_molecule}
In this section we extend our calculations to the treatment of 
the Ethylene molecule (C$_2$H$_4$) and show how these techniques permit to study
the time-dependence of molecular electronic states.
In particular we report on the clear observation of a strong $\pi\rightarrow\pi^*$ transition.

In order to have a good description of states close to the ionization threshold we employed
the asymptotically correct LB94 xc-functional in the adiabatic approximation~\cite{vanLeeuwen:1994bh}. 
We choose the molecular plane to be in the $xy$-plane with carbon atoms at coordinates $( \pm 1.26517 , 0, 0)$~a.u. and 
hydrogens at $(\pm 2.33230,  1.75518,0)$~a.u., $(\pm 2.33230, - 1.75518, 0)$~a.u.. 
The ion positions are held fixed during the time evolution. 
Norm-conserving Trouiller-Martin pesudopotentials are employed to describe  core electrons of 
Carbon. Moreover TDDFT equations are numerically integrated on a spherical grid with spacing $\Delta r=0.3$~a.u.,  
radius $R=16$~a.u. and 2~a.u. wide absorbing boundaries.

\begin{figure}
	\includegraphics[width=8cm]{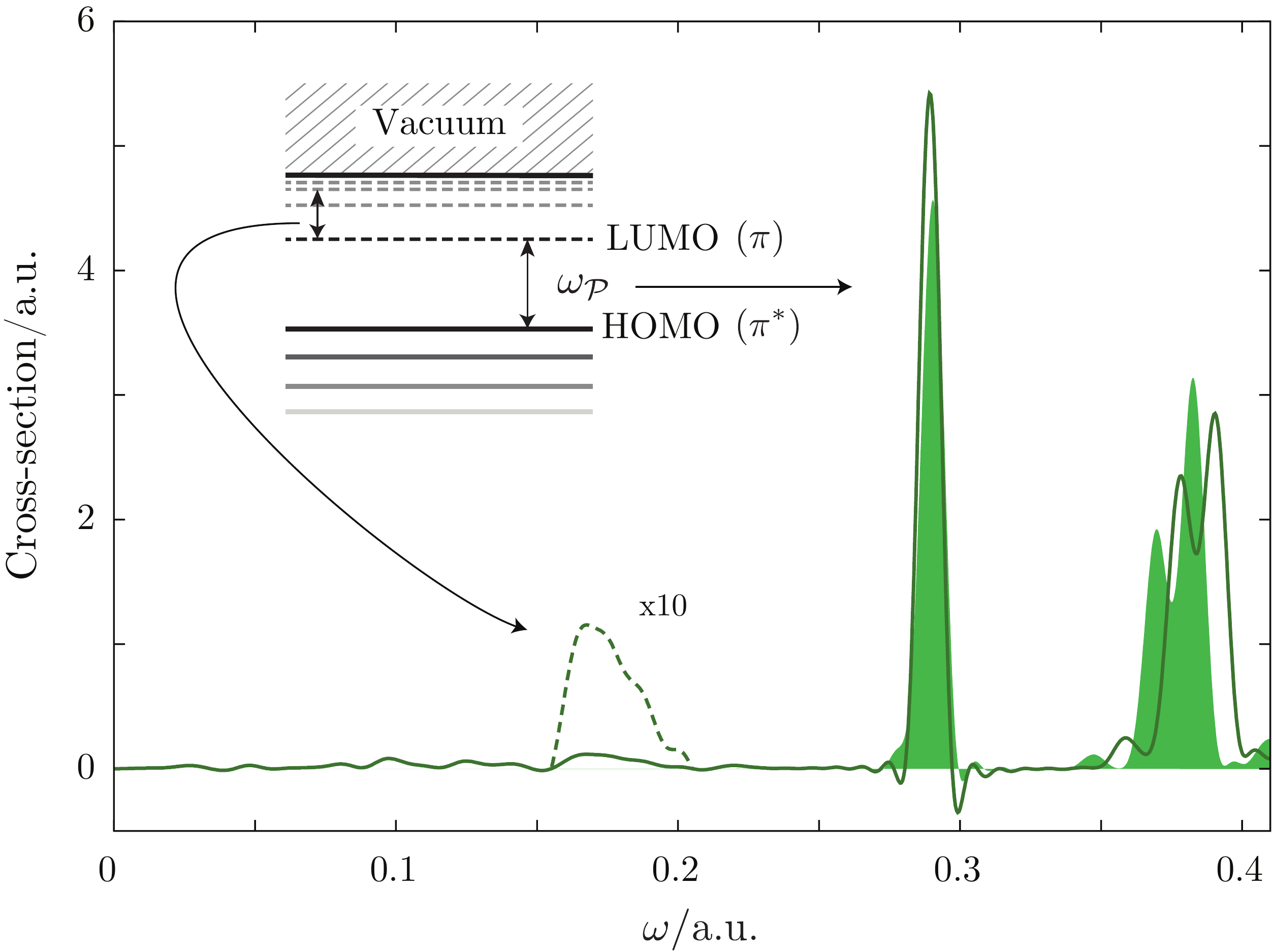}
	\caption{Comparison of the absorption spectra of unperturbed (filled curve) 
  and perturbed Ethylene molecule (solid line) below the ionization threshold.
  The molecule is excited by a 45 cycle $\sin^2$ envelope laser pulse polarized along the $x$-axis 
  with carrier $\omega_{\mathcal{P}}= 0.297$~a.u. of intensity $I=1.38\times10^{11}$ W/cm$^2$.}
	\label{fig:c2h4-abs-pp-cut}
\end{figure}
We perturb the system with a 15 cycles (3 cycles ramp) trapezoidal laser pulse
having carrier frequency $\omega_{\mathcal{P}}=0.297$~a.u. and intensity $I=1.38\times 10^{11}$~W/cm$^2$ polarized in the $x$-axis.
The laser frequency and the polarization direction are suited to excite mainly 
the molecular $\pi\rightarrow \pi^*$ transition.
The absorption spectrum of the excited molecule probed after the pulse Fig.~\ref{fig:c2h4-abs-pp-cut} shows  
the emergence of a peak associated with a population of the $\pi^*$ state.
Optical transitions from this excited state to high lying bound states
occur at energies lower than the HOMO-LUMO gap as illustrated by the scheme in Fig.~\ref{fig:c2h4-abs-pp-cut}.
Effects of the lack of memory in our xc-potential can be observed in the shifts of the 
peaks with respect to the known excitations of the unperturbed system. 
The characteristic excitations of a many-body system should not depend on 
the perturbation, unless we are in a strong light-matter coupling regime.

\begin{figure}
  \includegraphics[width=\columnwidth]{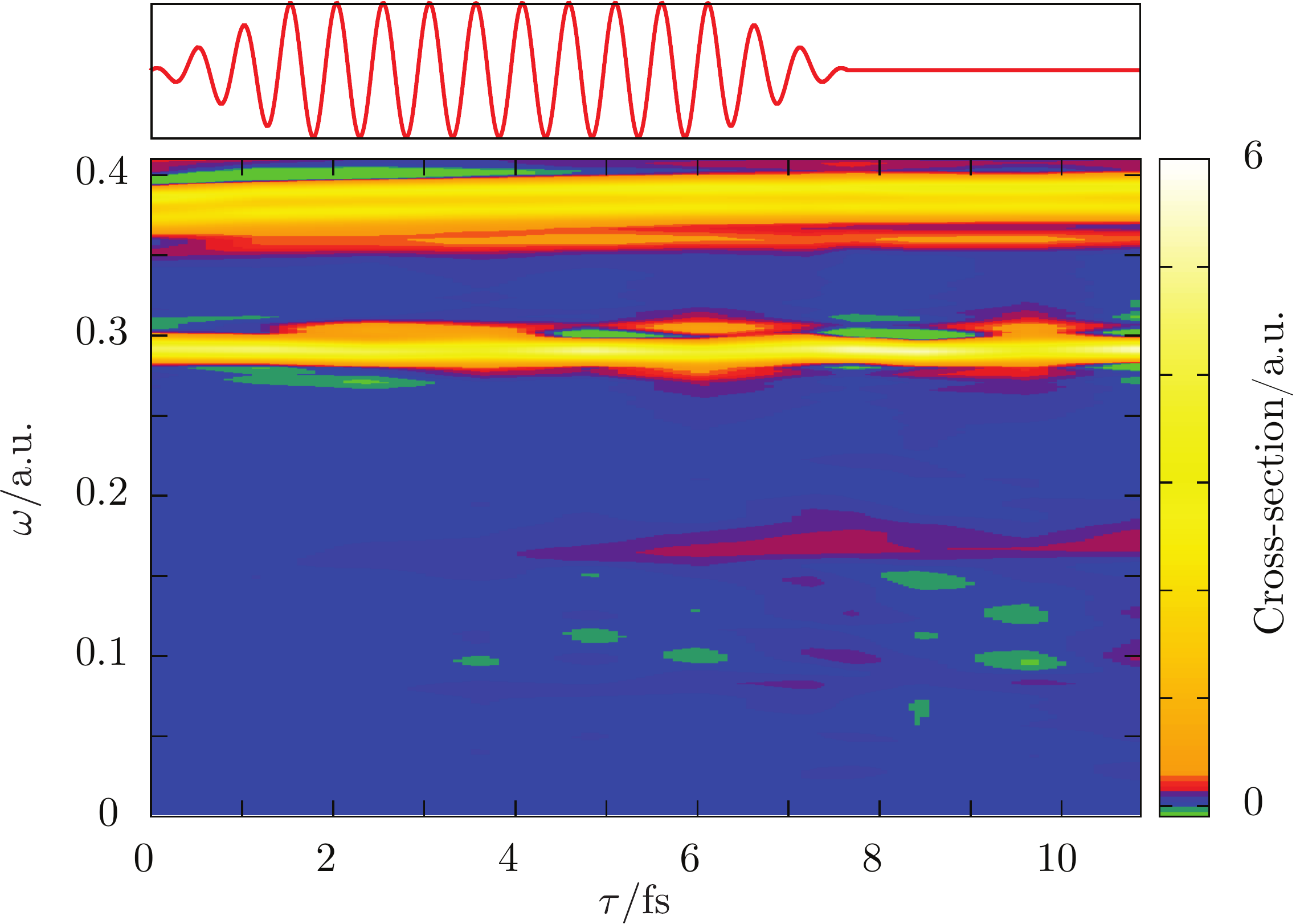}
	\caption{Ethylene molecule TAS. Pump laser pictured in 
  the upper panel is the same as in Fig.~\ref{fig:he-abs-pp-cut}.}
	\label{fig:c2h4_map-delay_time}  
\end{figure}
The build up of the transient spectrum as a function of time is shown in Fig.~\ref{fig:c2h4_map-delay_time}.
In comparison with the case of He discussed in Sec.\ref{sec:helium3d} the TAS map does not display any 
maxima during the pump time lapse due to the envelope area not crossing $\pi$ by the end of 
the pulse. A pulse with larger area would reveal the firs Rabi oscillation. 

\begin{figure}
    \includegraphics[width=\columnwidth]{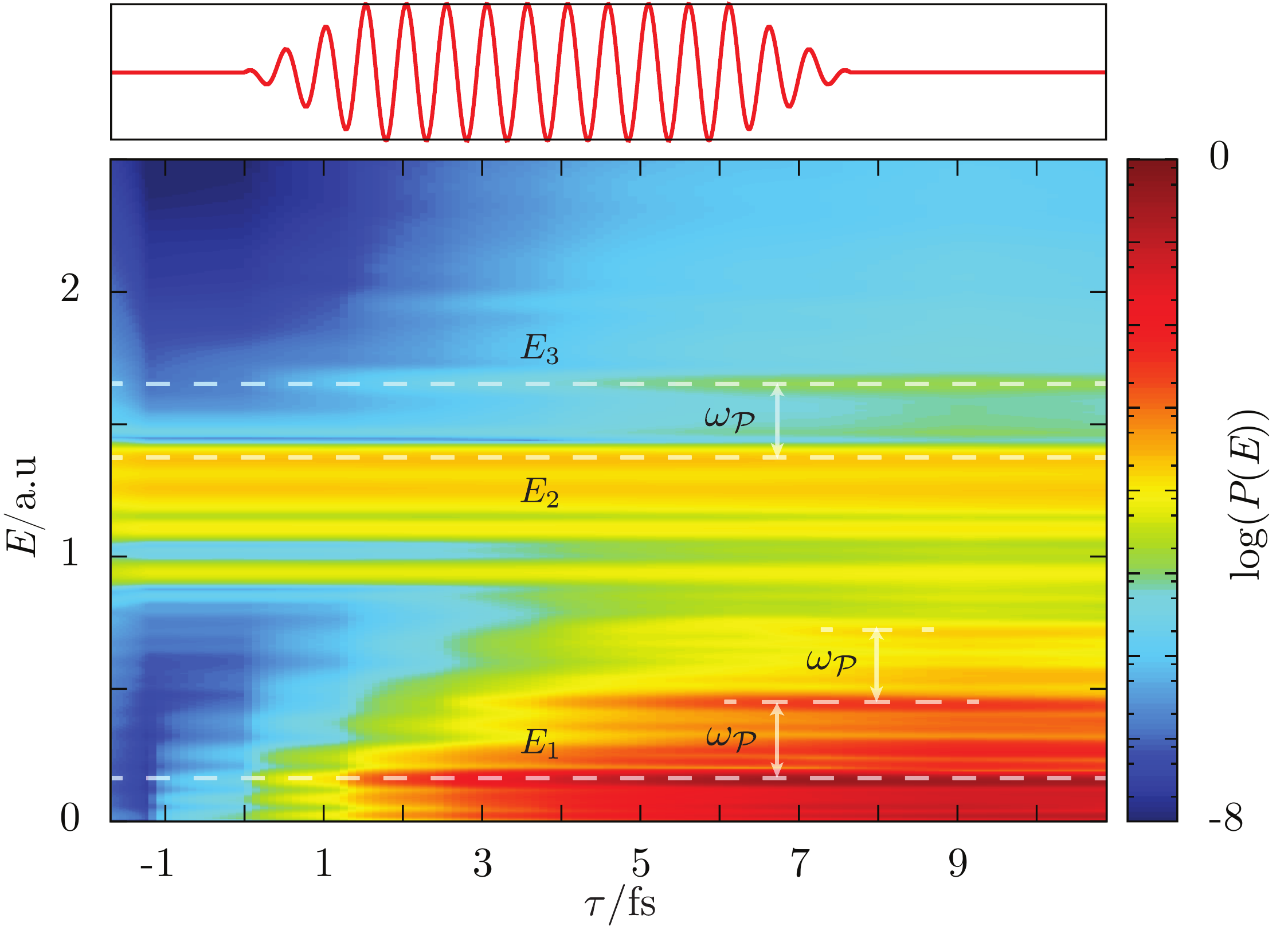}
    \caption{Logarithmic scale TRPES for C$_2$H$_4$. The molecule is probed at different delays with 
    is a 40 cycles trapezoidal laser pulse with 8 cycles ramp,  
    $\omega_p=1.8$ a.u., $I=5.4\times10^{9}$ W/cm$^2$ polarized along the $z$-axis perpendicular 
    with the pump. Pump laser (upper panel) is the same as in Fig.~\ref{fig:c2h4_map-delay_time}.  }
    \label{fig:C2H4_PES_pp}
\end{figure}
The TRPES map is presented in Fig.\ref{fig:C2H4_PES_pp}. Calculations were performed in a 
box of radius $R=30$~a.u. and the probe pulse is equal to the one used for the Helium atom in 
Sec.~\ref{sec:helium3d}, namely a 40 cycles (8 cycles ramp) trapezoidal pump at $\omega_p=1.8$~a.u. 
and $I=5.4\times 10^9$~W/cm$^2$, but polarized along the $z$-axis.
The choice of the polarization direction is important since, as we shall show,
the spectra may reveal geometrical features that depend on it.

\begin{figure}
    \includegraphics[width=\columnwidth]{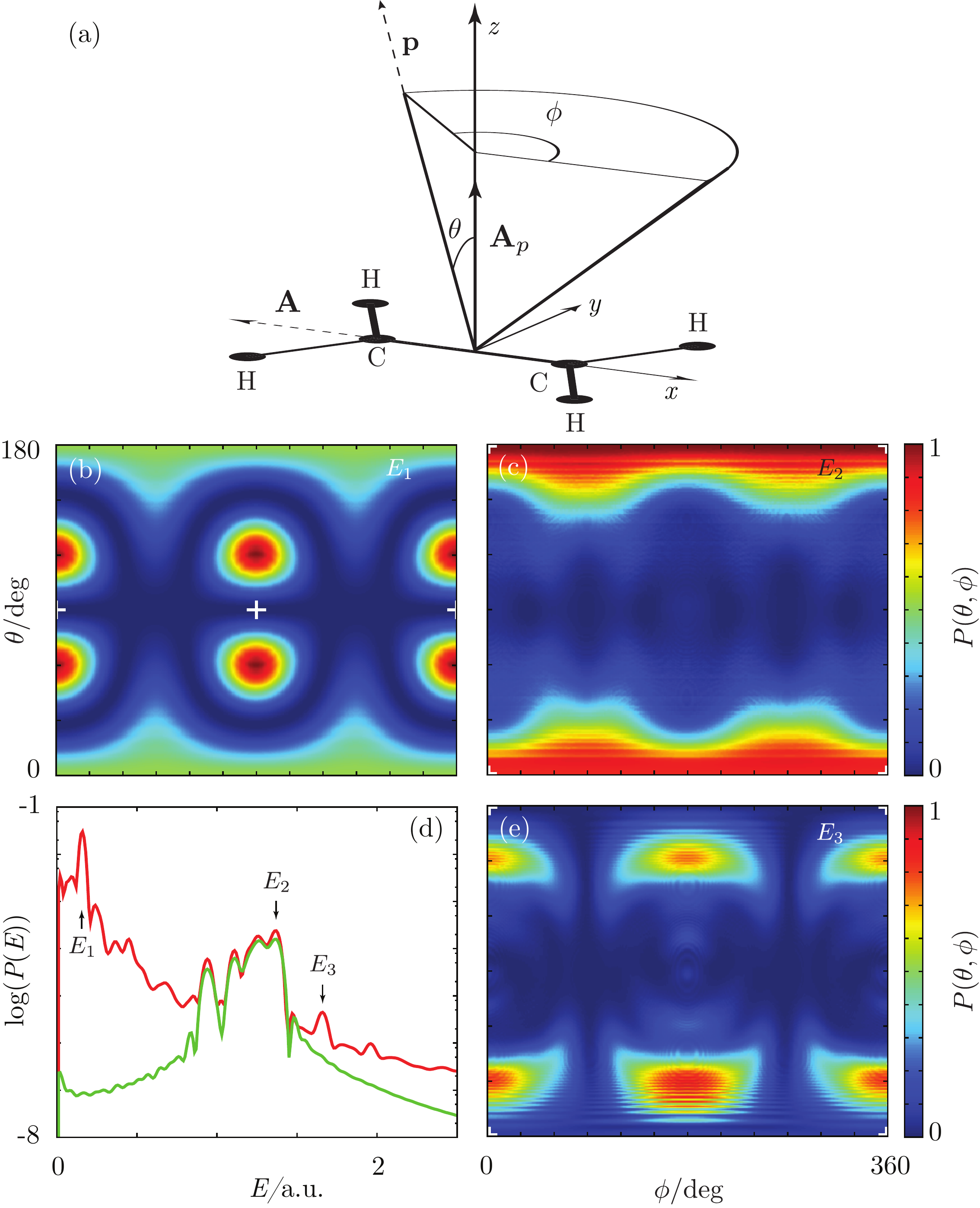}
    \caption{Angular- and energy- resolved photoelectron spectra for C$_2$H$_4$
    at two fixed delay times. Panel (a) geometry of the process: $\mathbf{p}$ indicates photoelectron 
    direction, $\mathbf{A}$ is the pump polarization vector, and $\mathbf{A}_p$ the probe one.
    Panel (d), logarithmic scale PES $P(E)$ for $\tau= -1.69$~fs (red) and  $\tau = 3.63$~fs (green). 
    Other panels depict normalized PADs $P(\theta,\phi, E)$ at $\tau = 3.63$~fs and energies 
    marked in (d): (b) $E_1=0.16$~a.u., (c) $E_2=1.37$~a.u., and (e) $E_3=1.67$~a.u..
    White marks indicate the position of the probe polarization vector (c), (e) 
    (at the corners) and the pump one (b) on the sphere. }
    \label{fig:C2H4_PES_panel}
\end{figure}

The TRPES displays a behavior similar to the one discussed in Sec.~\ref{sec:helium3d}.
A set of constant crests can be observed in the energy range from 0.8~a.u.to 1.5~a.u.. These are the 
peaks associated with electrons residing in the ground-state and ejected by 
the probe pulse. In particular the peak at $E_2=1.37$~a.u. corresponds to 
emission from the $\pi$ HOMO $E_2=\omega_p-\epsilon_{\pi}$, with $\epsilon_{\pi}=0.43$~a.u..
The $\pi$ orbital is localized on the Carbon atoms with two density lobes lying
in the $xz$-plane and nodes in the $xy$-plane (refer to Fig.~\ref{fig:C2H4_PES_panel} (a) for geometrical visualization).
A probe laser orientation along $y$ should suppress electrons on $xy$-plane perpendicular to its polarization and therefore
PAD $P(\theta,\phi, E_2)$ in Fig.~\ref{fig:C2H4_PES_panel} (c) for $\theta=0^{\circ}$ is
diminished  also due to geometrical reasons.
Signature of a $\pi$ symmetry can be clearly observed in the oscillations with $\phi$ presenting maxima at 
$\phi=90^{\circ}$, and $270^{\circ}$ along the plane perpendicular to $x$ that indicates 
a concentration along the $C-C$ bond axis, and minima for $\phi=0^{\circ},360^{\circ}$ consistent with 
a depletion in direction of each carbon atom.

Separated by a probe photon $\omega_{\mathcal{P}}$ at $E_3=\omega_p+\omega_{\mathcal{P}}-\epsilon_{\pi}=1.67$~a.u. 
we find photoelectrons ejected from the $\pi^*$ state. The intensity of the peak 
steadily increases in time accordingly to what is observed with TAS.
Compared with $\pi$, the $\pi^*$ orbital presents additional nodes on the
plane perpendicular to the molecular bond and a 
field polarized along $z$ is sensible to this kind of geometry.  
The PAD $P(\theta,\phi, E_3)$ in Fig.~\ref{fig:C2H4_PES_panel} (e) displays strong suppression
of electrons along the $yz$-plane at $\phi=90^{\circ},270^{\circ}$ and 
therefore presents a clear manifestation of photoemission from a $\pi^*$ state.

Slow electrons ejected at $E_1=2\omega_{\mathcal{P}}-\epsilon_{\pi}=0.16$~a.u. gradually 
increase and become the predominant ionization channel. 
The emergence in time of multi-photon peaks separated by $\omega_{\mathcal{P}}$ indicates that the pump 
is strong enough to trigger non-linear effects. 
These electrons are ejected after the simultaneous absorption of pump photons.
Electrons at $E_1$ reach the continuum with an $\omega_{\mathcal{P}}$ photon after the molecule 
has been excited to a $\pi^*$ state by another $\omega_{\mathcal{P}}$ photon.
PAD should therefore carry again signs of $\pi^*$ symmetry.    
It must be noted, that in this case, $\pi^*$ electrons are \emph{probed} with the 
pump itself, and therefore the laser polarization is along $x$.
As already discussed in the previous section, the laser polarization 
carries a geometrical factor of the form $\mathbf{A}\cdot\mathbf{p}$ with $\mathbf{A}=A \mathbf{x}$,
that introduces a suppression along the $yz$-plane ($\phi=90^{\circ},270^{\circ}$). 
Unfortunately, this plane is precisely where the $\pi^*$ photoemission minima should lay. 
For this reason PAD $P(\theta,\phi, E_3)$ in Fig.~\ref{fig:C2H4_PES_panel} (b) is not
suited to clearly discern a $\pi$ from a $\pi^*$ symmetry, and the suppression for $\theta=0^{\circ}$
along the $xy$-plane is compatible with both structures.


\section{Conclusions} 
\label{sec:conclusions}
In this work we studied the problem of describing ultrafast (attosecond scale) 
time-resolved absorption and photoemission 
in finite systems with TDDFT. We presented the theory and discussed how it can be implemented, such that
TDDFT can be successfully employed in the task of describing the dynamics of electronic excited states
in atoms and molecules. We illustrated the theory with three applications: 
one-dimensional Helium model, three-dimensional Helium atom, and Ethylene molecule. 

We studied the one-dimensional Helium atom perturbed by an external time dependent field exactly, 
by solving TDSE. We showed how it is possible to recover information about state populations 
through a comparison of the perturbed and unperturbed
absorption cross-sections, and that the population evolution in time can be described
in terms of Rabi physics. We then performed TDDFT calculations on the same model,
and we may conclude that the results obtained with the
EXX potential are in good agreement with the exact solution, although small artifacts appear
due to the wrong description of the functional memory dependence.

Furthermore we investigated the Helium atom in a more realistic three-dimensional treatment using the
EXX functional. 
We performed resonant pump-probe calculations monitoring both absorption and photoemission 
properties of the excited atom.
TAS turned out to be a sensible tool to monitor the build-up of the 
excited state, allowing to observe Rabi oscillations as a function of the time delay 
between pump and probe. 
TRPES also allowed the characterization of the excitation process in time. 
However, due to a dominant ionization channel associated with sequential two (pump) photons absorption, 
the information about the excited state population was less apparent.
Nonetheless PAD, being an observable sensitive to the geometrical arrangement of the 
ionized state, is a useful tool to discern the nature of each photoelectron 
peak.
As a final example we considered the case of the Ethylene molecule,
to study the time evolution of a $\pi\rightarrow\pi^*$ 
transition. PAD for ejected electrons offered clear evidence that the 
states taking part in the process were indeed of $\pi$ and $\pi^*$ nature. 

The theoretical framework that we have developed 
is a useful tool to understand and control non-equilibrium electronic dynamical processes in nanostructures 
and extended systems. 
New emergent properties of matter in the strong-coupling regime could appear 
that might give rise to new technological developments. 
Furthermore, monitoring electron and ion-dynamics provides fundamental 
insights into structure (i.e. time-resolved crystallography) 
and chemical processes in biology and materials science (e.g. for energy applications). 
There is plenty of room for new and fascinating discoveries about the 
behavior of matter under out-of-equilibrium conditions.

Still, from the fundamental point of view, there is a clear need for the development
of non-adibatic exchange and correlation functionals able to provide a reliable description 
of non-equilibrium processes and strong-light-matter interaction. 
Clearly, the methods presented here will automatically benefit from any 
theoretical advance in this direction. Conversely, the developers of new functionals
may take into account the correct description of pump-probe experiments as a useful
quality test.


\section{Acknowledgments} 
\label{sec:acknowledgments}
We acknowledge financial support from the European Research Council Advanced
Grant DYNamo (ERC-2010-AdG -Proposal No. 267374) Spanish Grants (FIS2011-65702-
C02-01 and PIB2010US-00652), ACI-Promociona (ACI2009-1036), Grupos Consolidados
UPV/EHU del Gobierno Vasco (IT-319-07), European Commission projects CRONOS
(280879-2 CRONOS CP-FP7),  THEMA(FP7-NMP-2008-SMALL-2, 228539), 
and CAPES Foundation, Ministry of Education of Brazil process number 2287/110. 
Computational time was granted by i2basque and BSC Red Espanola de Supercomputacion.



\end{document}